\documentclass[twocolumn]{aastex631}
\usepackage{bm}
\usepackage{comment}

\newcommand{\kms}{{\rm km\ s^{-1}}}

\begin{document}

\title{exoALMA XIV. Gas Surface Densities in the RX J1604.3-2130 A Disk from Pressure-broadened CO Line Wings}

\author[0000-0001-8002-8473]{Tomohiro C. Yoshida}
\correspondingauthor{Tomohiro C. Yoshida}
\email{tomohiroyoshida.astro@gmail.com}
\affiliation{National Astronomical Observatory of Japan, 2-21-1 Osawa, Mitaka, Tokyo 181-8588, Japan}
\affiliation{Department of Astronomical Science, The Graduate University for Advanced Studies, SOKENDAI, 2-21-1 Osawa, Mitaka, Tokyo 181-8588, Japan}

\author[0000-0003-2045-2154]{Pietro Curone} 
\affiliation{Dipartimento di Fisica, Universit\`a degli Studi di Milano, Via Celoria 16, 20133 Milano, Italy}
\affiliation{Departamento de Astronom\'ia, Universidad de Chile, Camino El Observatorio 1515, Las Condes, Santiago, Chile}

\author[0000-0002-0491-143X]{Jochen Stadler}
\affiliation{Laboratoire Lagrange, Université Côte d’Azur, CNRS, Observatoire de la Côte d’Azur, 06304 Nice, France}

\author[0000-0003-4689-2684]{Stefano Facchini}
\affiliation{Dipartimento di Fisica, Universit\`a degli Studi di Milano, Via Celoria 16, 20133 Milano, Italy}

\author[0000-0003-1534-5186]{Richard Teague}
\affiliation{Department of Earth, Atmospheric, and Planetary Sciences, Massachusetts Institute of Technology, Cambridge, MA 02139, USA}

\author[0000-0002-3001-0897]{Munetake Momose}
\affiliation{College of Science, Ibaraki University, 2-1-1 Bunkyo, Mito, Ibaraki 310-8512, Japan}


\author[0000-0003-2253-2270]{Sean M. Andrews}
\affiliation{Center for Astrophysics | Harvard \& Smithsonian, Cambridge, MA 02138, USA}

\author[0000-0001-7258-770X]{Jaehan Bae}
\affiliation{Department of Astronomy, University of Florida, Gainesville, FL 32611, USA}

\author[0000-0001-6378-7873]{Marcelo Barraza-Alfaro}
\affiliation{Department of Earth, Atmospheric, and Planetary Sciences, Massachusetts Institute of Technology, Cambridge, MA 02139, USA}

\author[0000-0002-7695-7605]{Myriam Benisty}
\affiliation{Laboratoire Lagrange, Université Côte d’Azur, CNRS, Observatoire de la Côte d’Azur, 06304 Nice, France}
\affiliation{Max-Planck Institute for Astronomy (MPIA), Königstuhl 17, 69117 Heidelberg, Germany}

\author[0000-0002-2700-9676]{Gianni Cataldi} 
\affiliation{National Astronomical Observatory of Japan, 2-21-1 Osawa, Mitaka, Tokyo 181-8588, Japan}


\author[0000-0003-4679-4072]{Daniele Fasano} 
\affiliation{Laboratoire Lagrange, Université Côte d’Azur, CNRS, Observatoire de la Côte d’Azur, 06304 Nice, France}

\author[0000-0002-9298-3029]{Mario Flock} 
\affiliation{Max-Planck Institute for Astronomy (MPIA), Königstuhl 17, 69117 Heidelberg, Germany}

\author[0000-0003-1117-9213]{Misato Fukagawa} 
\affiliation{National Astronomical Observatory of Japan, 2-21-1 Osawa, Mitaka, Tokyo 181-8588, Japan}

\author[0000-0002-5503-5476]{Maria Galloway-Sprietsma}
\affiliation{Department of Astronomy, University of Florida, Gainesville, FL 32611, USA}

\author[0000-0002-5910-4598]{Himanshi Garg}
\affiliation{School of Physics and Astronomy, Monash University, Clayton VIC 3800, Australia}

\author[0000-0002-8138-0425]{Cassandra Hall} 
\affiliation{Department of Physics and Astronomy, The University of Georgia, Athens, GA 30602, USA}
\affiliation{Center for Simulational Physics, The University of Georgia, Athens, GA 30602, USA}
\affiliation{Institute for Artificial Intelligence, The University of Georgia, Athens, GA, 30602, USA}

\author[0000-0001-6947-6072]{Jane Huang} 
\affiliation{Department of Astronomy, Columbia University, 538 W. 120th Street, Pupin Hall, New York, NY, USA}

\author[0000-0003-1008-1142]{John~D.~Ilee} 
\affiliation{School of Physics and Astronomy, University of Leeds, Leeds, UK, LS2 9JT}

\author[0000-0001-8446-3026]{Andr\'es F. Izquierdo} 
\affiliation{Department of Astronomy, University of Florida, Gainesville, FL 32611, USA}
\affiliation{Leiden Observatory, Leiden University, P.O. Box 9513, NL-2300 RA Leiden, The Netherlands}
\affiliation{European Southern Observatory, Karl-Schwarzschild-Str. 2, D-85748 Garching bei M\"unchen, Germany}
\altaffiliation{NASA Hubble Fellowship Program Sagan Fellow}

\author[0000-0001-7235-2417]{Kazuhiro Kanagawa} 
\affiliation{College of Science, Ibaraki University, 2-1-1 Bunkyo, Mito, Ibaraki 310-8512, Japan}

\author[0000-0002-8896-9435]{Geoffroy Lesur} 
\affiliation{Univ. Grenoble Alpes, CNRS, IPAG, 38000 Grenoble, France}

\author[0000-0003-4663-0318]{Cristiano Longarini} 
\affiliation{Institute of Astronomy, University of Cambridge, Madingley Road, CB3 0HA, Cambridge, UK}
\affiliation{Dipartimento di Fisica, Universit\`a degli Studi di Milano, Via Celoria 16, 20133 Milano, Italy}

\author[0000-0002-8932-1219]{Ryan A. Loomis}
\affiliation{National Radio Astronomy Observatory, 520 Edgemont Rd., Charlottesville, VA 22903, USA}


\author[0000-0003-4039-8933]{Ryuta Orihara} 
\affiliation{College of Science, Ibaraki University, 2-1-1 Bunkyo, Mito, Ibaraki 310-8512, Japan}

\author[0000-0001-5907-5179]{Christophe Pinte}
\affiliation{Univ. Grenoble Alpes, CNRS, IPAG, 38000 Grenoble, France}
\affiliation{School of Physics and Astronomy, Monash University, Clayton VIC 3800, Australia}

\author[0000-0002-4716-4235]{Daniel J. Price} 
\affiliation{School of Physics and Astronomy, Monash University, Clayton VIC 3800, Australia}

\author[0000-0003-4853-5736]{Giovanni Rosotti} 
\affiliation{Dipartimento di Fisica, Universit\`a degli Studi di Milano, Via Celoria 16, 20133 Milano, Italy}



\author[0000-0003-1412-893X]{Hsi-Wei Yen} 
\affiliation{Academia Sinica Institute of Astronomy \& Astrophysics, 11F of Astronomy-Mathematics Building, AS/NTU, No.1, Sec. 4, Roosevelt Rd, Taipei 10617, Taiwan}

\author[0000-0002-3468-9577]{Gaylor Wafflard-Fernandez} 
\affiliation{Univ. Grenoble Alpes, CNRS, IPAG, 38000 Grenoble, France}

\author[0000-0003-1526-7587	]{David J. Wilner} 
\affiliation{Center for Astrophysics | Harvard \& Smithsonian, Cambridge, MA 02138, USA}

\author[0000-0002-7501-9801]{Andrew J. Winter}
\affiliation{Laboratoire Lagrange, Université Côte d’Azur, CNRS, Observatoire de la Côte d’Azur, 06304 Nice, France}
\affiliation{Max-Planck Institute for Astronomy (MPIA), Königstuhl 17, 69117 Heidelberg, Germany}

\author[0000-0002-7212-2416]{Lisa W\"olfer} 
\affiliation{Department of Earth, Atmospheric, and Planetary Sciences, Massachusetts Institute of Technology, Cambridge, MA 02139, USA}

\author[0000-0001-9319-1296	]{Brianna Zawadzki} 
\affiliation{Department of Astronomy, Van Vleck Observatory, Wesleyan University, 96 Foss Hill Drive, Middletown, CT 06459, USA}
\affiliation{Department of Astronomy \& Astrophysics, 525 Davey Laboratory, The Pennsylvania State University, University Park, PA 16802, USA}





\begin{abstract}
The gas surface density is one of the most relevant physical quantities in protoplanetary disks.
However, { its precise measurement remains highly challenging due to the lack of a direct tracer.}
In this study, we report the spatially-resolved detection of pressure-broadened line wings in the CO $J=3-2$ line in the RX J1604.3-2130 A transition disk as part of the exoALMA large program.
Since pressure-broadened line wings are sensitive to the total gas { volume} density, we robustly constrain { the radial dependence of the gas surface density and midplane pressure in the region} located $50-110$ au from the central star, which encompasses the dust ring of the system.
The peak radius of the midplane pressure profile matches the dust ring { radial} location, directly proving radial dust trapping at a gas pressure maximum.
{ The peak gas surface density is $18-44\ {\rm g\ cm^{-2}}$} and decreases at radii interior to and exterior of the dust ring.
A comparison of the gas and dust surface densities suggests that the disk turbulence is as low as $\alpha_{\rm turb} \sim 2\times10^{-4}$.
{ Despite dust trapping}, the gas-to-dust surface density ratio at the ring peak is { $70-400$}, which implies already-formed protoplanets and/or less efficient dust trapping.
The gas surface density drop at radii interior to the ring is consistent with a gas gap induced by a Jupiter-mass planet.
The total gas mass within $50 < r < 110$ au is estimated to be $\sim 0.05-0.1\ M_\odot$ ($50-100\ {M_{\rm Jup}}$), suggesting that planetary system formation is possible.

\end{abstract}



\section{Introduction} \label{sec:intro}
Planetary systems are born in protoplanetary disks, where gas and dust grains orbit around a young star.
To understand the mechanisms of planet formation, it is crucial to observationally reveal the physical structure of these disks.
Since the Atacama Large Millimeter/sub-millimeter Array (ALMA) began operating, it has been revealed that substructures in dust continuum emission { in extended systems} are ubiquitous \citep[e.g.,][]{andr18}.
Among the different observed substructures, some of the most prominent ones are seen in transition disks, which often exhibit a mm-dust cavity of about 1-100 au and a radially confined dust ring.
It is believed that transition disks appear after the evolution of monolithic full-disks via several potential mechanisms such as grain growth in the inner region, disk clearing by planets, photoevaporation, and/or dust trapping at the outer edge of the magneto-rotational instability dead zone  \citep[e.g.,][]{vand23}.
Furthermore, in some cases, the radial accumulation of (sub) mm-sized dust grains may cause dust coagulation and could trigger the streaming instability, which is one of the most promising formation routes of terrestrial planets and cores of gas giant planets \citep[e.g.,][]{youd05, pini12}.

One of the most essential physical quantities to reveal the planet-forming potential in such rings is the gas surface density profile.
The gas in such regions is the main ingredient and as such, it sets the mass budget available for giant planet formation.
In addition, the gas affects the dynamics of dust grains.
Furthermore, gas dissipation after planet formation may affect the planets' orbits \citep[e.g.,][]{armi19}.
Therefore, it is crucial to measure the gas surface density profile.
However, despite its importance, measuring it is generally very challenging.
This is because the main component of the gas, H$_2$, does not efficiently emit radiation under typical conditions of protoplanetary disks due to the lack of a permanent dipole moment { and high upper-level energies for the rotational ladder}.
To trace the gas surface density profile, various indirect tracers such as dust thermal emission, optically thin CO isotopologue emission \citep[e.g.,][]{miot16}, and HD emission \citep[e.g.,][]{berg13}, have conventionally been used.
However, even for the total gas mass of one disk, the measured values are significantly scattered (a factor of $\sim10^3$) since the conversion factors { such as CO/H$_2$} are highly uncertain \citep{miot23}.
{ We note that the {\it relative} radial and vertical dependence of the gas pressure (and therefore density) structure can be constrained by analyzing the rotation curve of molecular lines \citep[e.g.,][]{teag18_hd163, teag18_as209, roso20, yu21, pezz25_163, Stadler_exoALMA}.}

Recently, \citet{yosh22} found very broad wings in the CO $J=3-2$ line in the inner region of the TW Hya disk and interpreted it as a result of pressure broadening.
Since the pressure-broadening occurs mainly due to collisions between CO and H$_2$ molecules, the pressure-broadened wings are sensitive to the total gas { volume and surface} densities.
They indeed constrain it without an assumption of the CO/H$_2$ ratio.

\begin{figure}[hbtp]
    \epsscale{1.0}
    \plotone{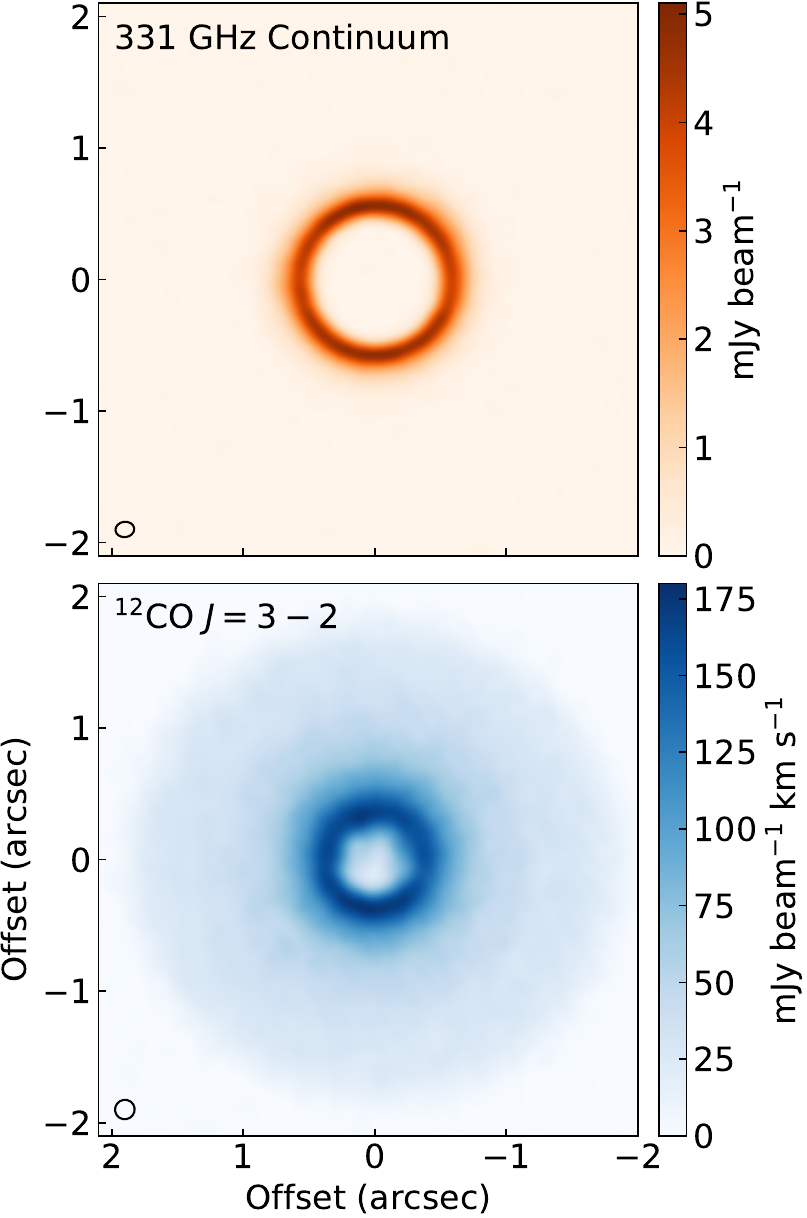}
    \caption{ (top) 331 GHz continuum image. (bottom) The integrated intensity map of the ${\rm ^{12}CO}\ J=3-2$ line. Both images are obtained as part of the exoALMA project \citep{Teague_exoALMA, Curone_exoALMA}. The J1604 disk is close to face-on and consists of a single well-defined continuum ring. }
    \label{fig:face}
\end{figure}
In this Letter, we focus on the transition disk around RX J1604.3-2130 A (hereafter J1604).
{ The central source is a K2 type pre-main sequence star \citep{luhm12} with a mass of $M_\star = 1.24\ M_\odot$ and age of $\sim 5-10$ Myr \citep{mana20}, which is located at 144.6 pc from the Earth \citep{gaia23} in the Upper Scorpius star-forming region \citep{deze99, prei99}.
The star is surrounded by a transition disk with a large central cavity
\citep[e.g.,][]{zhan14, maya18, vand21}.
The dust continuum emission consists of a single ring at a radius of 81 au from the central star (Figure \ref{fig:face}). The intensity of the ring sharply peaks just outside the cavity and steeply decreases toward the outside, suggesting dust trapping at the cavity edge \citep{stad23}.
This disk is one of the brightest disks in the Upper Scorpius star-forming region in terms of dust and gas emission \citep{math13, carp14, bare16}.
}
\citet{dong17} observed CO isotopologue lines as well as infrared and millimeter continuum emission and constrained the surface density structure via physical and chemical modeling.
{
They found that the gas surface density peaks at the dust ring and drops by three orders of magnitudes in the inner cavity.
The peak gas surface density at the ring and the total gas mass were estimated to be $\sim1\ {\rm g\ cm^{-2}}$ and $2.5\ M_{\rm Jup}$, respectively, although it relies on the potentially optically-thick $\rm C^{18}O$ line.
}
Inside and outside the dust cavity, \citet{stad23} detected a velocity deviation of CO gas from the Keplerian rotation, which can be caused by an embedded Jupiter-mass planet.
{ 
\citet{berg19} conducted observations of $\rm C_2H$, $\rm HCN$, and $\rm C^{18}O$ lines and detected all molecules, suggesting that the disk is generally molecular rich.
}
As part of the ALMA large program exoALMA \citep{Teague_exoALMA}, we observed the J1604 disk in the CO $J=3-2$ line with high angular and spectral resolution and sensitivity, which enables us to search { for faint pressure-broadened line wings}.
We note that this disk is the lowest-inclination disk among the exoALMA sample \citep{Curone_exoALMA}, which is preferable for the analysis as we can avoid the beam-smearing effect on the spectral shape.
{ Figure \ref{fig:face} shows the 331 GHz continuum and the integrated CO $J=3-2$ line images obtained as part of the exoALMA project.}

The structure of this Letter is the following.
{ In the next section, we describe our analysis of the dust continuum observations, which is necessary to interpret the gas observations. 
In Section \ref{sec:analysis_CO}, we report the wings detected in the $\rm ^{12}CO$ line and explain model fitting procedure.}
We present the results and discuss our interpretation in Section \ref{sec:res}.
We summarize our results in Section \ref{sec:sum}.

\section{{ Analysis of the dust emission}} \label{sec:analysis}
{ Before diving into the analysis of the CO emission line, which is the primary focus of this paper, we first present our modeling of the dust continuum disk and its corresponding results.
The continuum modeling is important for the spectral line fitting (Section \ref{sec:analysis_CO}).
Additionally, the derived dust and gas structures are compared in Section \ref{sec:res}.}

\subsection{{ SED Modeling}} \label{sec:cont}
We analyze the spectral energy distribution (SED) of multi-band dust continuum observations to retrieve the dust disk properties such as the dust surface density profile, maximum dust size, and temperature profile.
We use continuum images previously observed with ALMA at 1.3 mm (Band 6) and 3.1 mm (Band 3) in addition to the new exoALMA continuum image at 0.9 mm \citep[Band 7, { a beam size of $0\farcs14\times0\farcs12$}; ][]{Curone_exoALMA}.
For Band 6, we use the image published by \citet{stad23}, which has a beam size of $0\farcs060 \times 0\farcs039$.
We newly analyze archival observations (\# 2015.1.00819.S; PI L.\ Ricci) for the Band 3 image, { which has a beam size of $2\farcs7 \times 1\farcs7$}. The reduction processes of the Band 3 { data} are described in Appendix \ref{sec:b3im}.
{ The CLEAN images are re-convolved to have circularized beams of $2\farcs7,\ 0\farcs061$,\ and $0\farcs14$, for Band 3, 6, and 7, respectively.}

The beam sizes are different in the three wavelengths; the Band 6 image clearly resolves the narrow ring structure, but the Band 3 image does not even resolve the whole disk.
Therefore, we perform simple parametric modeling to retrieve the dust disk properties.
Since the J1604 disk consists of a single ring with a shoulder on the outer side of the ring \citep{maya18, stad23}, we assume that the dust surface density $\Sigma_d$ can be expressed by superposition of two Gaussian rings, that is,
\begin{eqnarray}
\label{eq:sigmad}
    \Sigma_d(r) = \Sigma_{d, 1} \exp\left[ -\frac{1}{2} \left( \frac{r-r_{d, 1}}{ \sigma_{d, 1}} \right)^2 \right] \\ + \Sigma_{d, 2} \exp\left[ -\frac{1}{2} \left( \frac{r-r_{d, 2}}{ \sigma_{d, 2}} \right)^2 \right],
\end{eqnarray}
where $\Sigma_{d, i}$, $r_{d, i}$, and $\sigma_{d, i}$ represent the peak dust surface density, ring radius, and ring width for the ring index $i=(1,2)$, respectively.
The midplane temperature profile is assumed to be a power law, 
\begin{equation}
\label{eq:tmid}
    T_{\rm mid}(r) = T_{\rm 100} ( r/100 {\rm au} ) ^ {-0.5}.
\end{equation}
In addition, we assume that the dust size distribution is homogeneous { in the ring.
We set a grain size distribution as a power law with a fixed index of $-3.5$, and a flexible maximum grain size of $a_{\rm max}$.
Note that the minimum grain size is fixed to $10^{-5}$ cm, but this assumption does not affect the results.
}
The dust opacities at each wavelength are calculated using {\tt dsharp\_opac}, assuming the DSHARP dust properties \citep{birn18}.

The dust emission for each location of the disk is then calculated in the same manner as in recent similar studies \citep[e.g.,][]{sier21, maci21, guid22} based on a solution of the radiative transfer equation when considering self-scattering by dust grains \citep{miya93, liu19, zhu19}.
First, we calculate the intensity of a dust slab at each wavelength as a function of radius according to Equation (11) of \citet{zhu19}.
Then, we assume the geometry of the J1604 disk { \citep[$i = 9.5^\circ,\ {\rm PA} = 124.9^\circ$, consistent with][]{Curone_exoALMA}} and distribute the intensity on the 2D image plane.
Finally, we convolve the images with { the beams} and resample the image with intervals of the beam size to properly treat the noise correlation in a beam.
Our dust emission model has eight free parameters; six parameters for the surface density profile in Equation (\ref{eq:sigmad}), $T_{\rm 100}$, and $\log_{10} a_{\rm max}$.

The likelihood function of the model images is calculated by
\begin{equation}
    \chi ^2 = \sum_\nu \bm{R_\nu}^{T} C_\nu^{-1} \bm{R_\nu},
\end{equation}
where $\bm{R_\nu}$ is the residual vector between the modeled and observed image at a frequency $\nu$.
Note that the observed images are also re-sampled in the same manner as the model images.
$C_\nu$ is the variance-covariance matrix at $\nu$.
The summation is taken for three frequencies.
In general, the absolute flux error dominates the uncertainty in the case of SED analysis.
This uncertainty can be treated as a spatially correlated noise over the image assuming that the flux uncertainty follows a known probability distribution.
We adopt the following formulation for an $(i, j)$ element of $C$;
\begin{equation}
    C_{i, j} = \sigma^2 \delta_{i, j} + s_f^2 M_i M_j.
\end{equation}
where $\delta_{i, j}$ is the Kronecker delta, $\sigma$ is the image noise level, $s_f$ is the flux scaling uncertainty (i.e. $s_f = 0.1$ for a flux uncertainty of $10\%$), and $M_i$ is the $i$-th element of the model image.

In practice, we set $s_f = 0$ for all bands first and sampled the posterior probability distribution by the Markov-Chain Monte-Carlo (MCMC) method using {\tt emcee} \citep{fore13}.
We adopt 5000 steps with 24 walkers for our MCMC chain.
After confirming convergence, we calculate the median values from the last 4200 steps.
The model images created by the median values are used for $M_i$ and $M_j$.
Then, we allow $s_f$ to be non-zero to estimate the uncertainty of each parameter.
While the ALMA technical handbook states the flux accuracy for flux calibrators is 2.5\% for Band 3 and 5\% for Band 6 and 7, the resultant uncertainty on the image plane could be larger due to e.g., phase error and imaging processes.
According to \citet{fran20} and \citet{yosh25}, we conservatively assume that the resultant uncertainty is three times larger than the nominal value, substitute $s_f = [ 0.075, 0.15, 0.15 ]$ for Band 3, 6, and 7, respectively, and perform the second round of MCMC sampling.
In the second round, the uncertainties of some parameters became larger as expected.
We compare the best-fit model and observations without pixel re-sampling in Appendix \ref{sec:cont_comparison}.
The relative difference between observations and models is almost within the flux uncertainties.

\subsection{ Results of the SED modeling}\label{sec:cont_res}
\begin{table}[hbtp]
\caption{Best-fit parameters of the multi-band continuum modeling. { The uncertainties are 16th and 84th percentiles of the posterior probability distributions.}}
 \label{tab:dust}
  \begin{tabular}{lcc}
   \hline \hline
   Parameter & Best-fit value & Unit \\
   \hline
   $r_{d, 1}$ & $82.81^{+0.01}_{-0.01}$ & au \\
   $\log_{10} \Sigma_{d, 1}$ & $-0.50^{+0.04}_{-0.03}$ & $\rm g\ cm^{-2}$ \\
   $\sigma_{d, 1}$ & $3.09^{+0.03}_{-0.04}$ & au \\
   \hline
   $r_{d, 2}$ & $96.65^{+0.09}_{-0.08}$ & au \\
   $\log_{10} \Sigma_{d, 2}$ & $-1.31^{+0.02}_{-0.02}$ & $\rm g\ cm^{-2}$ \\
   $\sigma_{d, 2}$ & $15.29^{+0.05}_{-0.05}$ & au \\
   \hline
   $T_{100}$ & $14^{+1}_{-1}$ & K \\
   $\log_{10} a_{\rm max}$ & $-0.85^{+0.05}_{-0.03}$ & cm \\
   \hline
  \end{tabular}
\end{table}

The simple model { described in Section \ref{sec:cont}} can reproduce the observed images and the parameters are well constrained.
The best-fit parameters of the multi-band continuum modeling are presented in Table \ref{tab:dust}.
The resulting dust surface density profile is plotted in Figure \ref{fig:sigmas_and_T}(a).
The peak dust surface density is estimated to be $0.35^{+0.03}_{-0.02}\ {\rm g\ cm^{-2}}$ at a radius of $\simeq 82$ au, which is in a typical range found in other (bright) protoplanetary disks \citep[e.g., ][]{maci21, guid22, sier21}.
The midplane temperature at $r = 82$ and $100$ au is constrained to be $15\pm1\ {\rm K}$ and $14\pm 1\ {\rm K}$, respectively.
The maximum dust size $a_{\rm amax}$ is estimated to be $1.4^{+0.2}_{-0.1}\ {\rm mm}$.
Although we assume that $a_{\rm amax}$ is uniform across the ring, there is no strong residual in all images (Figure \ref{fig:cont}), suggesting the assumption is reasonable.
From the dust surface density profile, we estimate a total dust mass of $2.7^{+0.2}_{-0.1}\times10^{-4}\ {M_\odot}$.

We also test a model with the temperature radial slope being $-0.4$ instead of $-0.5$.
While the best-fit temperature increases by $\sim10\%$, the peak dust surface density decreases by $\sim10\%$.
{ We note that the assumption of the power-law temperature profile might be not very accurate, given that the temperature structure is no longer radially smooth in transition disks \citep[e.g.,][]{clee11, brud13}.
Higher spatial resolution observations (especially in Band 3) are needed to more robustly determine the dust disk properties}.

\section{{ Analysis of the CO emission}} \label{sec:analysis_CO}
{ In this Section, we explain how we analyze the CO line data. 
The results are presented in Section \ref{sec:res}.
}

\subsection{Stacked spectra and visual inspection} \label{sec:stacked}
In the following analysis of the CO line, we use one of the fiducial exoALMA image cubes of the CO $J=3-2$ line \citep{Loomis_exoALMA}.
The synthesized beam is circularized and has a full width at half maximum (FWHM)  of $0\farcs15$.
The channel width is set to $0.1\ \kms$.

\begin{figure*}[hbtp]
    \epsscale{1.1}
    \plotone{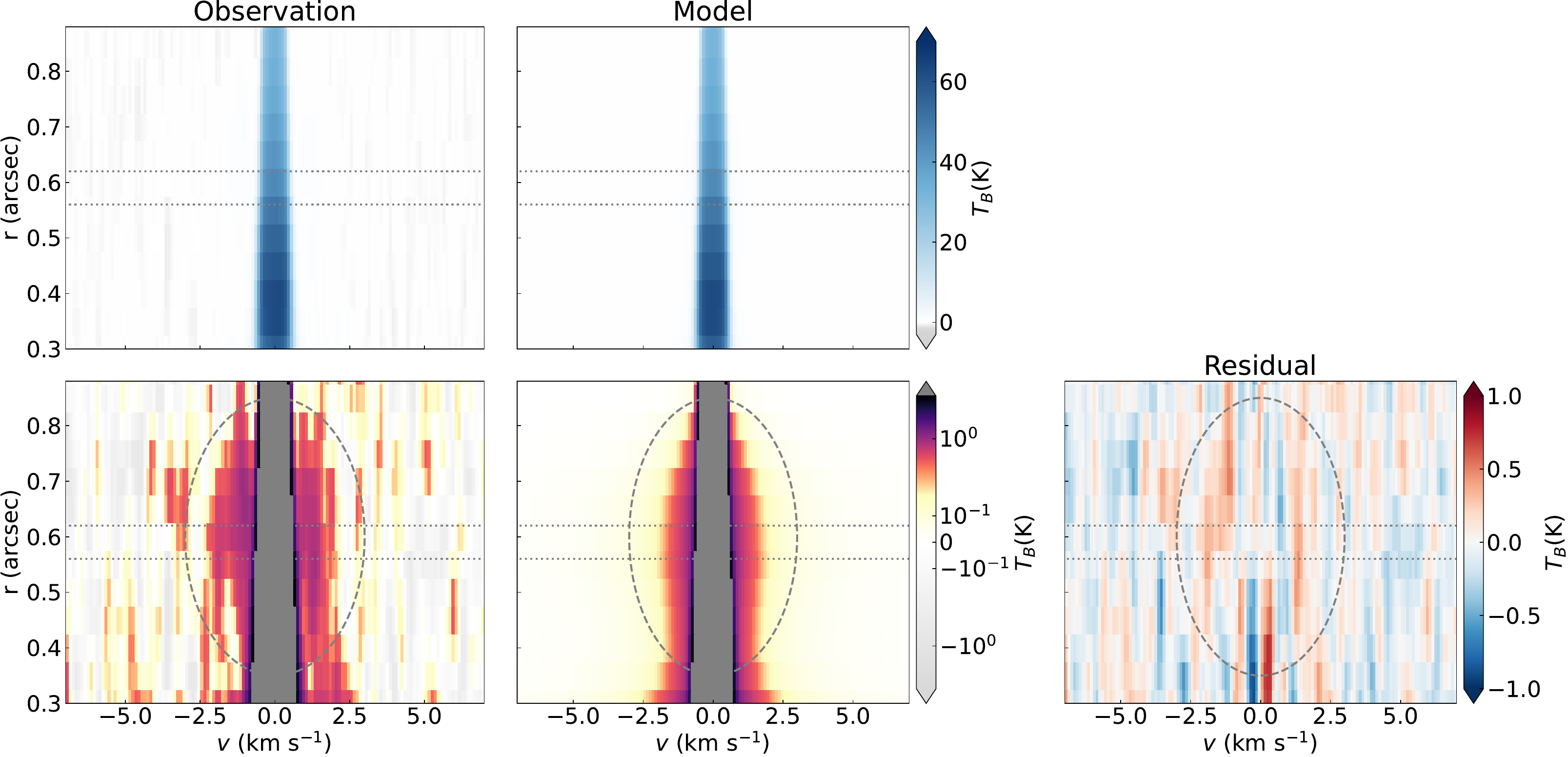}
    \caption{ (left) Azimuthally stacked CO spectra. The upper and lower panels show the same data, { but in the lower panel, we convolved the spectrum at each radius with a flat kernel of three channels and use the symmetric logarithm color map in order to enhance the faint emission in the line wings.} The horizontal dotted lines indicate the { radial range} of the dust ring. The dashed ellipse marks the { faint line wings we analyze in this Letter}. (middle) The best-fit model of the CO spectra. (right) The residual between the observed and model spectra.}
    \label{fig:teardrop}
\end{figure*}
To search for the faint wing feature due to pressure broadening, we need to obtain a high signal-to-noise ratio, and therefore, we azimuthally-stack the spectra.
First, we create the line-of-sight velocity map by fitting a tapered Gaussian with an optically thick core to the observed spectrum at each position using {\tt bettermoments} \citep{teag18}.
Then, the velocity map is fitted by the geometrically thin Keplerian rotation model using {\tt eddy} \citep{eddy} to determine the position angle, stellar mass, and central position of the disk.
The inclination angle is fixed to $i = 6\fdg0$ \citep{stad23}.
We note that a possible error in the inclination angle does not affect the results (Appendix \ref{sec:azimuthal}).
Adopting the best-fit parameters, which are consistent with results based on {\tt discminer} by \citet{Izquierdo_exoALMA}, we pick up an annulus with a radial width of $0\farcs05$ and fit the velocity variation as a function of the azimuthal angle by
\begin{equation}
v_{\rm LOS} = v_\phi \cos(\phi) \sin(i) - v_r \sin(\phi) \sin(i) + v_{\rm LSR},
\end{equation}
where $v_{\rm LOS}$ and $\phi$ are the line-of-sight velocity and azimuthal angle on the disk coordinate.
$v_{\rm LSR}$ denotes the systemic velocity but contains any vertical motion within the disk as well.
Fitting parameters, $v_\phi$ and $v_r$, represent the azimuthal and radial velocity, respectively.
We use the module {\tt eddy.linecube} to do this fitting and repeated the procedure for all radii.
The original spectra are then shifted by the derived line-of-sight velocities.
In Appendix \ref{sec:azimuthal} (Figure \ref{fig:stacking}), we present the shifted spectra as a function of radius, azimuthal angle, and velocity. It is clear that the shifting works well.
Finally, we average the spectra along the azimuthal axis and obtain the azimuthally stacked CO spectra as shown in Figure \ref{fig:teardrop}.
When stacking the spectra, { we make the assumption} that the line wings are identical along the azimuthal direction.
We have checked that there is no significant azimuthal dependence (Appendix \ref{sec:azimuthal}).

In the stacked spectra, the peak intensity { presents} its maximum at $\sim 0\farcs4$, which is consistent with \citet{stad23} and \citet{Galloway_exoALMA}.
Interestingly, the spectra have relatively broad line wings at $r=0\farcs4-0\farcs8$.
\begin{figure}[hbtp]
    \epsscale{0.9}
    \plotone{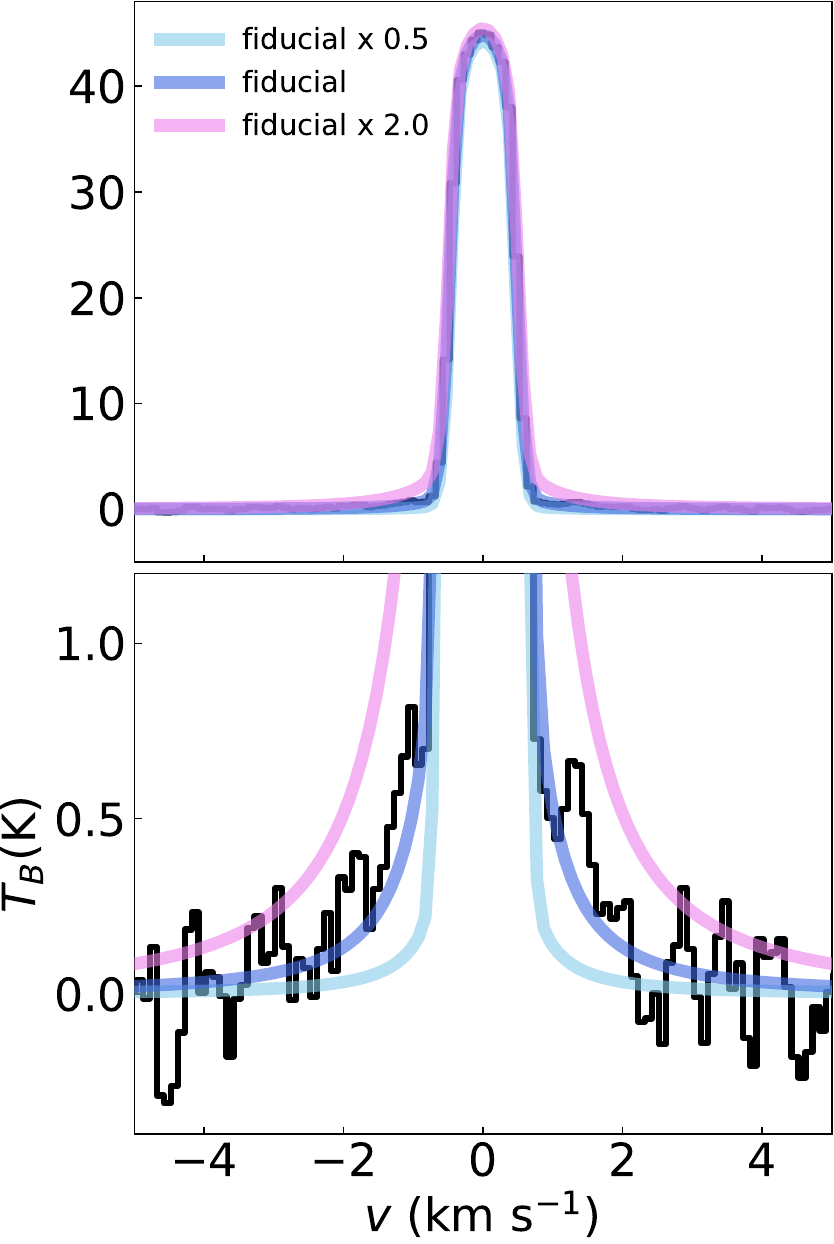}
    \caption{ Radially averaged spectrum at $r=0\farcs4-0\farcs8$ (see Figure \ref{fig:teardrop}). The lower panel shows the zoom-in view of the upper panel. Colored lines indicate models with different gas surface densities as discussed in Section \ref{sec:gas_res} }
    \label{fig:spec}
\end{figure}
The azimuthally stacked spectra are radially averaged at radii between $r=0\farcs4$ and $r=0\farcs8$, and are plotted in Figure \ref{fig:spec}, which shows the wing feature.
\begin{figure}[hbtp]
    \epsscale{0.9}
    \plotone{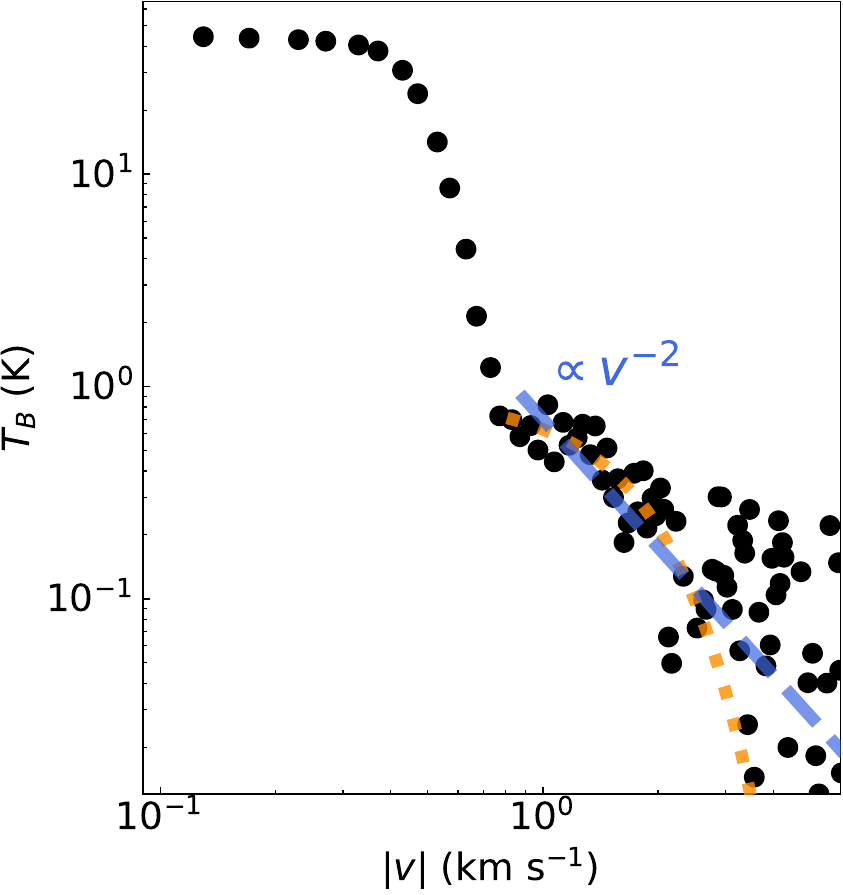}
    \caption{ Log-log plot of the radially averaged spectrum at $r=0\farcs4-0\farcs8$ as a function of $|v|$. The wing ($|v| > 1\ \kms$) emission follows $v^{-2}$, which is predicted for pressure-broadened line wings. The orange dotted line indicates the best-fit model if we interpret the wing as a high-temperature component. As discussed in Section \ref{sec:stacked}, the derived temperature from the line width is $4.8\times10^{3}\ {\rm K}$, which is unrealistically high for CO molecules.}
    \label{fig:lorenzian}
\end{figure}
In Figure \ref{fig:lorenzian}, the averaged spectrum is plotted against the absolute velocity on a log-log scale.
The wing emission at $|v| \gtrsim 1\ \kms$ is almost proportional to $v^{-2}$.
If an optical depth profile follows the Gaussian function, the line wings should damp with $\propto \exp(-v^2/2\sigma)$, where $\sigma$ is the line width.
Therefore, the observed line profile implies a non-Gaussian line-broadening process.

To quantitatively test if the observed line wing can be explained only by Gaussian broadening, we model the wing emission by an isothermal slab,
\begin{equation}
    I = B(T)(1-e^{-\tau}), \label{eq:toymodel}\\
\end{equation}
where
\begin{equation}
    \tau = \tau_{0} \exp \left( -\frac{m_{\rm CO} v^2}{2 k_B T} \right),
\end{equation}
$B$ is the Planck function, $T$ is the temperature, $\tau_{0}$ denotes the optical depth at the line center, $m_{\rm CO}$ and $k_B$ are the molecular mass of CO and Boltzmann constant, respectively.
We fitted the high-velocity range ($|v| > 0.8\ \kms$) of the averaged spectrum by Equation (\ref{eq:toymodel}) using {\tt scipy.optimize.curve\_fit}, after converting the units to brightness temperature.
The best-fit parameters are $T \simeq 4.8\times10^3\ {\rm K}$ and $\tau_{0} \simeq 1.9 \times10^{-4}$.
The best-fit spectrum is also plotted with the orange dotted line in Figure \ref{fig:lorenzian}, which matches the observations at small $v$ but deviates at larger $v$.
The best-fit temperature of $T \simeq 4.8\times10^3\ {\rm K}$ might suggest that the emission { originates} from the uppermost layer of the disk.
However, since CO molecules are thermally { dissociated} at $\sim2000$ K, such a high temperature is unphysical.
Therefore, the observed broad line wings cannot be explained by a high-temperature component.
We note that sub-sonic turbulent broadening \citep[e.g.,][]{flah18} is also unlikely as a cause of the observed wings because such broadening effects should be substantially smaller than thermal broadening, which means that the very high temperature is still needed even if turbulence exists.
In addition to turbulence, we note that any microscopic non-Keplerian motion can be excluded as long as it is sub-sonic.

Larger scale non-Keplerian motion of the CO gas could also enhance the line width.
However, the velocity deviation from the Keplerian rotation does not exceed $\sim 0.1\ \kms$ at these radii \citep{stad23, Teague_exoALMA}, which is much smaller than the observed width.

Another potential scenario is a disk wind that is vertically mirror symmetric against the disk midplane.
However, the continuum modeling (Section \ref{sec:cont}) suggests that the optical depth of the $\sim80$ au continuum ring is $\sim5$ in Band 7.
Therefore, the emission from the redshifted side of the disk wind should be absorbed by the continuum disk.
The observed symmetric spectrum shows that the disk winds cannot explain the wings.

A candidate of the function that has a damping wing $\propto v^{-2}$ is the Lorentzian function, or its convolution with Gaussian, the Voigt function.
The most promising mechanism that makes the line profile Voigt-like is pressure broadening \citep[e.g.,][]{radpro, yosh22}.
Pressure broadening is a general process of line broadening in astrophysics.
When a CO molecule collides with another molecule, mainly H$_2$ or He, the radiation field is disrupted.
If the collision randomly occurs by following the Poisson probability distribution as a function of time, the line profile becomes the Lorentzian function.
Since the line profile without collision is the Gaussian function, the resultant line profile is the convolution of those two functions, the Voigt function.
Indeed, \citet{yosh22} found broad line wings in the TW Hya disk and reproduced the observed spectrum by taking pressure broadening into account.
Importantly, the intensity of the pressure-broadened line wings depends on the gas volume density near the midplane, which is challenging to trace with other methods.

\subsection{Modeling the spectra} \label{sec:mod}
To derive the gas surface density profile, we model the azimuthally stacked spectra after considering pressure broadening.
First, we assume the vertical temperature structure.
We used a parametric function that describes a layered temperature structure \citep[e.g.,][]{dart03, andr12, law22},
\begin{eqnarray}
    T(z) &=& T_0 \\
    &+& 
    \left\{
    \begin{array}{ll}
    T_{\rm mid} & (|z| < z_1) \\
    \frac{T_{\rm atm} - T_{\rm mid}}{2} \cos\left( \frac{\pi}{2} \frac{|z| - z_1}{z_2 - z_1} \right) & ( z_1 \leq |z| < z_2 ) \\
    T_{\rm atm} & (z_2 \leq |z|)
    \end{array}
    \right.
\end{eqnarray}
Here $z$ is the height from the disk midplane.
$T_{\rm atm}$ represents the temperature in the atmosphere and $z_1$ and $z_2$ are the characteristic heights of transition from the midplane to transition layer and transition layer to the atmosphere, respectively.
$T_{\rm mid}$ is given in Equation (\ref{eq:tmid}).
We note that $z_1$ and $z_2$ are practically given as follows:
\begin{eqnarray}
    z_1 &=& f_{z_1} H_g \\
    z_2 &=& f_{z_2} H_g.
\end{eqnarray}
Here, $f_{z_1}$ and $f_{z_w} \equiv f_{z_2}-f_{z_1}$ are free parameters in the fitting.
$H_g$ is the gas scale height and is given by
\begin{equation}
    H_g = \sqrt{ \frac{k_B T_{\rm mid} r^3}{\mu m_p G M_\star} },
\end{equation}
where $G$ is the gravitational constant and $m_p$ denotes the proton mass.
$\mu$ is the mean molecular mass and is set to $2.37$.
Additionally, we introduced 
\begin{eqnarray}
    T_0 = 
    \left\{
    \begin{array}{ll}
    0\ & (|z| < z_1) \\
    p \frac{|z| - z_1}{z_2}\ & (|z| \geq z_1)
    \end{array}
    \right.
\end{eqnarray}
The background temperature gradient in addition to the two-layer structure with arctangent-like transition has been seen in radiative transfer calculations \citep{dull02}.
Indeed, this parameter is needed to fit the core component of the spectra based on our tests.

{ In our model fitting procedure, we fix $T_{\rm 100} = 14$ K as a fiducial case. This is not only based on our continuum modeling in Section \ref{sec:cont_res} but also a typical value around $r=100$ au in a T Tauri disk without a cavity \citep[e.g.,][]{dull02}.
However, considering that the dust disk modeling presents its own uncertainties (Section \ref{sec:cont_res}), this value may be underestimated for three reasons;
First, the detection of the pressure-broadened line wings suggests that the gaseous CO molecules survive around the disk midplane below one gas scale height, where the gas density is high.
However, since the CO sublimation temperature is generally thought to be $\sim20$ K \citep[e.g.,][]{mini22}, the CO molecules may be frozen near the midplane, although we note that observations imply some diversity depending on disks \citep[$\sim13-28$ K; ][]{qi13, schw16, zhan17, pint18, qi19}.
Second, it is theoretically predicted that the midplane temperature rises to $\sim30$ K around a cavity edge of a transition disk \citep{clee11, brud13}, which should be in similar conditions to the J1604 disk.
Third, a disk survey by \citet{berg19} suggests that the J1604 disk has the highest excitation temperatures of the $\rm HCN$ and $\rm C_2H$ lines among their samples that include Herbig disks (see also Appendix \ref{app:c18o}).
Although these lines may trace relatively elevated layers \citep[e.g.,][]{pane23}, it is likely that the midplane temperature is also relatively higher than typical T Tauri disks.
Therefore, we also consider another case with $T_{\rm 100} = 30$ K as an upper side of the midplane temperature.
}

The H$_2$ gas density structure as a function of height, $n_{\rm H_2}(z)$, is determined by assuming the vertical hydrostatic equilibrium self-consistent with the temperature structure \citep{Longarini_exoALMA}.
The CO density structure is written as
\begin{equation}
    n_{\rm CO}(z) = X_{\rm CO} n_{\rm H_2}(z),
\end{equation}
where $X_{\rm CO}$ is the CO/H$_2$ abundance ratio.
We tentatively assume $X_{\rm CO}=10^{-4}$ for fitting and discuss possible uncertainties in the following sections.

We integrate the radiative transfer equation using the above CO density and temperature structures, and then, create a synthetic spectrum.
Here, we also take the continuum absorption into account.
In practice, we first calculate the line emission from the back side of the disk.
Then, we multiply it by a factor of $\exp({-\tau_d})$, where $\tau_d$ is the corresponding optical depth of the continuum emission at 346 GHz derived by the multi-band continuum modeling.
Finally, we add the continuum intensity and restart the integration of the radiative transfer equation for the front side to take the potential absorption by the continuum disk into account.
The continuum emission is then subtracted from the spectra for a fair comparison.

For the line opacity profile, we adopt the Voigt function and follow \citet{yosh22}.
For the spectroscopic constants and broadening parameters, we use the LAMDA database \citep{scho05, yang10} and HITRAN database \citep{HITRAN22, tan22}, respectively.
We approximate that the (bulk) velocity of CO gas is constant along the line of sight at each position since the disk is nearly face-on \citep{lank23}.
In addition, we convolve the spectrum with a Gaussian that represents the effective velocity resolution.
Note that the velocity resolution is limited not only by the instrumental resolution but also by the intensity gradient in a beam \citep{kepp19, pezz25_163}.
Finally, we shift the spectrum along the velocity axis by $\Delta v$ and interpolate at the observed velocity channel.

The convolved synthetic spectrum at each radius $r$ is then compared with the azimuthally stacked spectrum at the same radius.
Here, there are seven free parameters, which are the gas surface density $\log_{\rm 10} \Sigma_g$, difference between atmospheric temperature and midplane temperature $\Delta T \equiv T_{\rm atm}-T_{\rm mid}$, parameters on the temperature structure, $f_{z_1}$, $f_{z_w}$, and $p$, velocity shift $\Delta v$, and the effective velocity resolution $\sigma_v$.
We sample the posterior probability distribution of the model spectrum at each radius using the MCMC method with {\tt emcee} \citep{fore13}.
We set uniform priors for all parameters with boundaries of $\log_{\rm 10} \Sigma_g = [-3, 3]\ {\rm g\ cm^{-2}}$, $\Delta T = [ 0, 2000 ]\ {\rm K}$, $f_{z_1} = [0, 50]$, $f_{z_w} = [0, 50]$, $p = [0, 50]$, $\Delta v = [-0.1, 0.1]\ {\rm km\ s^{-1}}$, and $\sigma_v = [0.02, 1.0]\ {\rm km\ s^{-1}}$.
For each radius, our MCMC chain consists of 2000 steps with 32 walkers.
We run the sampling process, discard the first 1000 steps, and calculate the median values.
Then, we initialize the chains again using the median values.
We re-run MCMC sampling for 6000 steps and check that the chains are well converged.
Finally, after discarding the first 3000 steps, the median values of the chain are adopted as fiducial parameters.
We present the typical marginalized posterior distribution in Appendix \ref{app:corner} (Figure \ref{fig:corner}).

Furthermore, to check if pressure-broadening is indeed required, we turn off the Voigt function but use the Gaussian function as the opacity profile and perform the same fitting procedure as in the above case.
The resultant best-fit model and its residual are plotted in Appendix \ref{sec:gaussian}.
We confirm that the spectra cannot be well-fitted without pressure-broadening.

\section{Results and Discussion}\label{sec:res}

\subsection{Gas surface density profile}\label{sec:gas_res}
We compare the best-fit CO spectrum at $r = 0\farcs4-0\farcs8$ { for the fiducial temperature ($T_{\rm 100} = 14$ K)} with observations in Figure \ref{fig:spec}.
{ Note that the best-fit spectrum for the high-temperature case with $T_{\rm 100} = 30$ K is identical to the above case.}
If we reduce the gas surface density by a factor of 2, the observed wings are not reproduced. On the other hand, if we scale the gas surface density by 2, the wings are much brighter than the observations (color lines in Figure \ref{fig:spec}).
{ Clearly, the broad line wings are extremely sensitive to the gas surface density.
This is because the optical depth in the wing is proportional to the squared gas volume density and the wings are optically thin \citep{yosh22}.
}
We discuss the CO emitting layer as a function of $v$ in Appendix \ref{app:layer}. 
The two-dimensional temperature structure simultaneously constrained by the model fitting is discussed in Appendix \ref{app:temp}.

\begin{figure*}[hbtp]
    \epsscale{1.1}
    \plotone{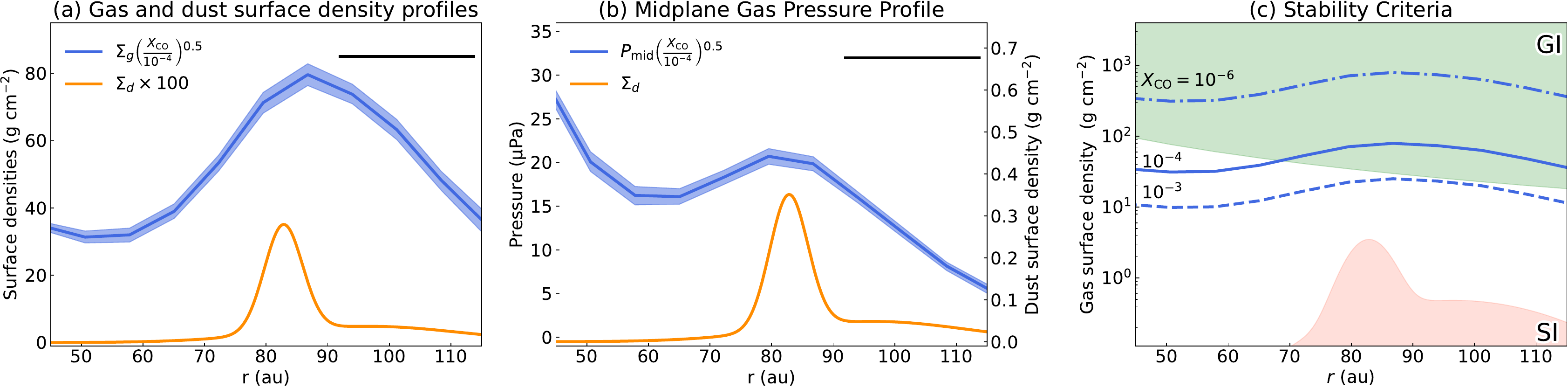}
    \caption{ { Fitting results for the fiducial temperature assumption.} (a) Gas (blue) and dust (orange) surface density profiles. The dust surface density is multiplied by 100 for visibility purposes. The black bar at the upper right corner indicates the spatial resolution of the CO image. (b) Midplane gas pressure profile (blue) with an arbitrarily scaled dust surface density profile (orange). The dust surface density peaks at the gas pressure maximum. 
    Note that there are uncertainties from the undetermined $X_{\rm CO}$ in panels (a) and (b).
    (c) Gas surface density profile for various $X_{\rm CO}$ with criteria for gravitational instability and streaming instability. }
    \label{fig:sigmas_and_T}
\end{figure*}

\begin{figure*}[hbtp]
    \epsscale{1.1}
    \plotone{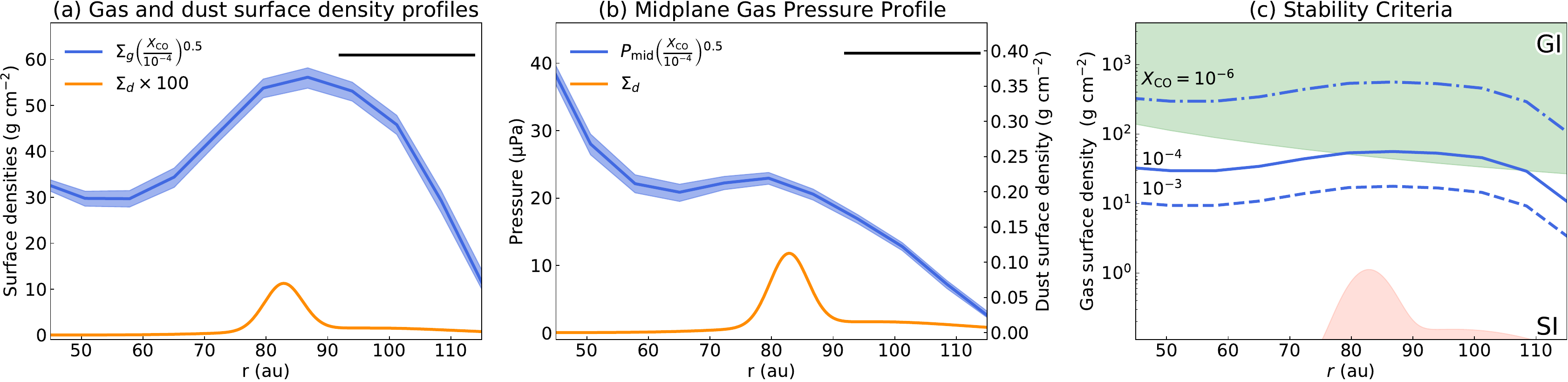}
    \caption{ { Same as Figure \ref{fig:sigmas_and_T}, but for the high midplane temperature case with $T_{\rm 100} = 30$ K. } }
    \label{fig:sigmas_and_T_highT}
\end{figure*}

{ We plot the best-fit gas and dust surface density profiles in Figure \ref{fig:sigmas_and_T} (a) and \ref{fig:sigmas_and_T_highT} (a) for the fiducial and higher temperature cases, respectively.
In both cases, the gas surface density decreases at the radii within and outside the dust ring.
The maximum value at $r\sim87$ au is $\sim80 (X_{\rm CO}/10^{-4})^{-0.5}\ {\rm g\ cm^{-2}}$ and $\sim56 (X_{\rm CO}/10^{-4})^{-0.5}\ {\rm g\ cm^{-2}}$ for the fiducial and high temperature cases, respectively.
}
A factor of $(X_{\rm CO}/10^{-4})^{-0.5}$ is included here because we assume $X_{\rm CO} = 10^{-4}$ in the model fitting, and the intensity in the pressure-broadened line wings depends on the product of the squared gas surface density and the CO/H$_2$ ratio \citep{yosh22}.
This factor can be reduced by measuring the column density of CO via observations of optically thin CO isotopologue lines such as $\rm ^{13}C^{18}O$ or $\rm ^{13}C^{17}O$.
As of writing this Letter, there is no observation of these isotopologues in the J1604 disk.
The $\rm ^{13}CO$ line observed by the exoALMA project is significantly optically thick \citep{Galloway_exoALMA}.
{ We also check the $\rm C^{18}O\ {\rm J=2-1}$ line in the ALMA Science Archive and the peak brightness temperature around the ring exceeds $\sim30$ K in the full Planck function.
This suggests that the $\rm C^{18}O$ line is also highly optically thick (Appendix \ref{app:c18o}).
}
{
It is also notable that we can eliminate the uncertainty from the temperature assumption by observing an optically thin isotopologue line because such lines should have the same source function as the line wing of $\rm ^{12}CO$ and the ratio of them determines the gas density \citep{yosh22}.
}
Future observations will allow us to robustly determine the absolute value of the gas surface density profile.

{ However, there are some constraints on $X_{\rm CO}$ \citep{ober23}.}
According to detailed chemical network calculations with dust evolution by \citet{krij18, krij20}, the gas phase $X_{\rm CO}$ can vary from $10^{-3}$ to $10^{-6}$ in protoplanetary disks.
The lower value can result from the locking of gas phase CO to large dust grains and their radial drift.
On the other hand, the higher value may arise interior to the CO snowline due to transportation and sublimation of the CO ice from the outer disk.
Observations by \citet{zhan20} also suggest that $X_{\rm CO}$ decreases from $10^{-4}$ to $10^{-6}$ as a function of age.
If we take a conservative range of $10^{-6} < X_{\rm CO} < 10^{-3}$, the remaining factor in the gas surface density ranges from $0.3 - 10$.
Therefore, we can reduce the uncertainty by a factor of $\sim30$ compared to conventional gas surface density measurements using CO isotopologue lines that are proportional to $X_{\rm CO}$.

We calculate Toomre Q \citep{toom64} from the inferred gas surface density profiles.
{ In the case of the fiducial temperature assumption,} Toomre Q is estimated to be lower than unity at $r>70$ au if $X_{\rm CO} = 10^{-4}$, which might indicate that the disk is gravitationally unstable.
However, there is no significant asymmetric structure such as a spiral arm in the J1604 disk, suggesting the disk is gravitationally stable and $Q > 1$ in reality.
If we assume $Q > 1$ at the peak of the gas surface density, we can obtain $X_{\rm CO} > 7 \times 10^{-4}$.
{ By combining the conservative range of $X_{\rm CO}$, it is suggested that the CO is not depleted around the ring, with $7 \times 10^{-4} < X_{\rm CO} < 10^{-3}$.
Even the lower limit is higher than the elemental abundance of carbon in the interstellar medium, $\rm C/H \simeq 3.2\times10^{-4}$ \citep{hens21}, implying an additional supply of carbon by the radial drift and trapping of dust grains \citep{krij18, zhan20_co}.
}
{ Adopting the range of $X_{\rm CO}$, the peak gas surface density is constrained to be $25-30\ {\rm g\ cm^{-2}}$.}
{ On the other hand, assuming the high midplane temperature, the peak gas surface density is constrained to be $18-44\ {\rm g\ cm^{-2}}$ by similarly assuming that the disk is gravitationally stable.
In addition, $X_{\rm CO}$ can be estimated to be $1.6 \times 10^{-4} < X_{\rm CO} < 10^{-3}$ in this case.
The lower side becomes consistent with the interstellar abundance.
Considering both temperature assumptions, we conclude the peak gas surface density of $18-44\ {\rm g\ cm^{-2}}$ with $1.6 \times 10^{-4} < X_{\rm CO} < 10^{-3}$.
}

{ We can also estimate the gas mass within $50<r<110$ au of $0.05-0.1\ {\rm M_\odot}$ (or $50-100\ {\rm M_{\rm Jup}}$) by integrating the derived gas surface density profile and assuming $X_{\rm CO}$ is constant over this radial range.
The gas mass within the continuum ring ($75<r<100$ au) is also estimated to be $\sim0.02-0.06\ {\rm M_\odot}$ (or $\sim 20-60\ {\rm M_{\rm Jup}}$).
As the mass is much larger than several Jupiter masses, we conclude that there is enough gas for planetary system formation in the J1604 disk.}
{ 
The gas surface density and mass is at least $\sim20$ times larger than the estimate by \citet{dong17}.
This descrepancy may be explained by the optically thick $\rm C^{18}O$ line (Appendix \ref{app:c18o}) which was effectively used as a mass tracer in \citet{dong17}.
}

We also show the midplane gas pressure profile,
\begin{equation}
    P_{\rm mid} = n_{\rm H_2}(0) k_B T,
\end{equation}
in Figure \ref{fig:sigmas_and_T}(b) and \ref{fig:sigmas_and_T_highT}(b) for the different temperature cases.
{ The $P_{\rm mid}$ profile peaks at $r\sim80$ au with $\sim21(X_{\rm CO}/10^{-4})^{-0.5}\ {\rm \mu Pa}$ and $\sim 23(X_{\rm CO}/10^{-4})^{-0.5}\ {\rm \mu Pa}$ for the fiducial and higher temperature cases, respectively, which matches the peak of the dust surface density.}
This is direct evidence that the dust grains are radially trapped at the pressure maximum.
In addition, there is no detection of a significant sub-structure in the gas and dust surface density beyond the peak.
Therefore, it is suggested that dust grains radially drifted from the outer region towards the ring.
However, we estimate the gas-to-dust surface density ratio at the { dust surface density peak to be $70-400$. Here, the dust surface density for the higher temperature case is estimated by multiplying $B(T=14\ {\rm K})/B(T = 30\ {\rm K})$, where $B(T)$ is the Planck function, to the fiducial case to compensate the temperature difference.}
This gas-to-dust surface density ratio is broadly similar to { or larger than} that in the interstellar medium.
In the case of dust trapping in gas pressure maxima, however, significant dust enrichment by a factor of 10-100 is predicted \citep[e.g.,][]{birn10}.
The moderate or no dust enrichment in dust rings could be explained by two scenarios.
First, it is possible that small dust grains can still be advected across the ring, suppressing the dust enrichment.
Second, there might be significant hidden masses, such as planetesimals, that cannot be observed at millimeter wavelengths \citep{stam19}.

The gas surface density drops by a factor of $\sim 2-2.5$ at $r\sim60$ au compared to the ring peak.
{ This is consistent with \citet{Stadler_exoALMA}, who found evidence of gas density cavity in the inner region by analyzing the rotation curves of $\rm ^{12}CO$ and $\rm ^{13}CO$. }
This gas drop could be interpreted as a gap induced by a giant planet.
Indeed, \citet{stad23} suggests that the velocity deviation in this disk can be explained by a giant planet with a mass of $1.6-2.9\ M_{\rm jup}$ at $r=41 \pm 10$ au.
Assuming that the background gas surface density at $r\sim60$ au { (before carved by the planet)} is $\sim 110 (X_{\rm CO}/10^{-4})^{-0.5}\ {\rm g\ cm^{-2}}$ by a rough extrapolation from the outer region, the depth of the gas gap is estimated to be $\sim 0.3$.
{ According to \citet{kana15}, we can estimate the mass of the planet of $0.1-2\ M_{\rm jup}$ for a range of turbulence parameter $\alpha = 10^{-4}-10^{-2}$, which is also roughly consistent with the results of \citet{stad23}.
Note that \citet{stad23} fixed the disk aspect ratio at the planet location to 0.1 while we used self-consistent values. If we adopt the disk aspect ratio of 0.1 instead, the mass of the planet is estimated to be $0.3-3\ M_{\rm jup}$.
}

\subsection{Turbulence and dust fragmentation}
The degree of accumulation of the dust grains at the gas pressure maximum depends on the turbulence strength.
Therefore, with the derived gas and dust surface density profiles, we can estimate the disk turbulence strength \citep{dull18, roso20}.
The gas surface density profile can be related to the dust surface density profile by the advection-diffusion equation.
We can write its steady-state solution as
\begin{equation}
\Sigma_d = \epsilon_0 \Sigma_g \exp\left[  \int_{r_0}^{r} \frac{St}{\alpha_{\rm turb}} \frac{d \ln{P_{\rm mid}}}{dr} dr \right],
\end{equation}
where $\epsilon_0$ is the dust-to-gas surface density ratio at an arbitrary radius $r_0$, and $\alpha_{\rm turb}$ is the turbulence strength \citep{sier19}.  
$St$ is the Stokes parameter and is given by
\begin{equation}
St = \frac{\pi \rho_m a_{\rm rep}}{2 \Sigma_g}.
\end{equation}
Here, $\rho_m$ is the dust material density.
We introduce a representative dust radius $a_{\rm rep}$, to which the observations are most sensitive, rather than the maximum dust size.
We adopt $a_{\rm rep} = 0.6\ {\rm mm}$ by converting $a_{\rm max} = 1.4\ {\rm mm}$, which is derived from the multi-band continuum modeling, using Figure 10 of \citet{doi23}.

\begin{figure}[hbtp]
    \epsscale{0.9}
    \plotone{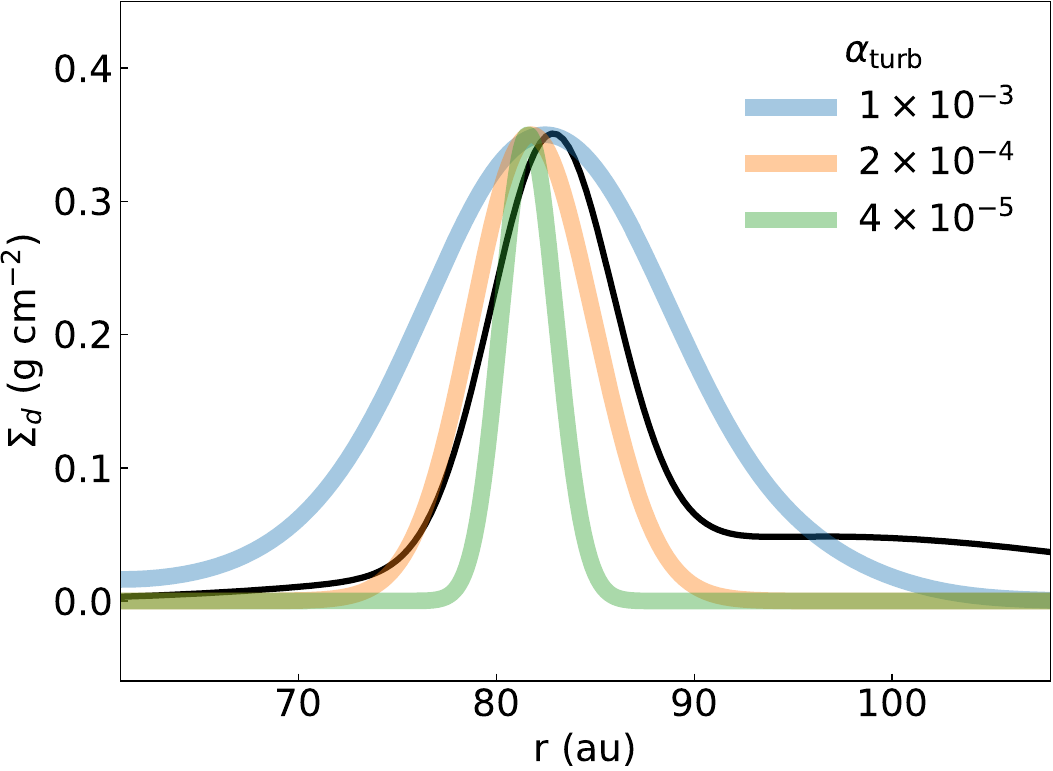}
    \caption{ Observed dust surface density profile (black) and calculated profiles from the observed gas surface density for different $\alpha_{\rm turb}$. }
    \label{fig:turb}
\end{figure}

Using the observed gas surface density and midplane pressure profiles { for the fiducial temperature case with $X_{\rm CO} = 7 \times 10^{-4}$}, we calculate $\Sigma_d$ for $\alpha_{\rm turb} = [1\times10^{-3}, 2\times10^{-4}, 4\times10^{-5}]$ and compared them with the observed dust surface density profile in Figure \ref{fig:turb}.
Note that the peak dust surface densities are always normalized to $0.35 {\rm\ g\ cm^{-2}}$ as in the observations.
When $\alpha_{\rm turb}$ is set to $2\times10^{-4}$, the observed width of the dust ring is well reproduced.
Such a weak level of turbulence near the midplane is also inferred by multiple observations \citep[e.g.][]{flah18, flah20, vill22, yosh22} and might be common among disks.
{ Note that assumption of the midplane temperature does not affect the widths significantly although the predicted dust peak slightly shifts towards the disk center for the higher temperature case. }

With the turbulence strength and Stokes parameter, we can estimate the dust fragmentation threshold velocity $v_{\rm frag}$.
Assuming that the maximum grain size is determined by the turbulence-induced collisional fragmentation, $v_{\rm frag}$ is expressed as
\begin{equation}
v_{\rm frag} = \sqrt{3 St_{\rm max} \alpha_{\rm turb}} c_s,
\end{equation}
where $St_{\rm max}$ is the Stokes parameter for the maximum dust size $a_{\rm max}$, $c_s$ is the sound speed \citep[e.g.,][]{birn09, ueda19, ueda20, zaga23}.
Using the constrained values, $v_{\rm frag}$ is estimated to be { $\sim 1\ {\rm m\ s^{-1}}$ at the dust ring. }
This low fragmentation velocity { is roughly consistent with recent observational results \citep{jian24, ueda24, yosh25} } and may be explained by grains covered by less sticky material \citep{musi16, okuz19, arak21}.
{ Another possibility is that the maximum dust size can be limited by the bouncing, not the fragmentation, which also may explain the small dust size \citep[e.g.,][]{domin24}. }
We note that this estimate also depends on the accuracy of $a_{\rm max}$, which could be improved by high-resolution observations in longer wavelengths.

\subsection{Comparison with other targets}
As part of the exoALMA large program, 
{ several independent attempts were made to measure disk masses: \citet{Longarini_exoALMA} by using rotation curves to determine dynamical masses; \citet{Trapman_exoALMA} by forward-modeling the $\rm N_2H^{+}$ and $\rm C^{18}O$; finally, \citet{Rosotti_exoALMA} by leveraging the CO emission height as a proxy for disk mass.}
Unfortunately, the J1604 disk is excluded from their sample because of its low inclination which makes kinematical measurements difficult.
However, it may be interesting to compare the J1604 disk with other exoALMA targets \citep{Teague_exoALMA}.

The estimated mass { within $50<r<110$ au} of the J1604 disk is a lower limit of the total mass since the CO emission is detected up to $r\sim265$ au \citep{stad23} { while the contribution from the inner cavity may be negligible as the $\rm ^{12}CO$ emission is not strong \citep{Teague_exoALMA}.}
{ Assuming that the gas surface density is $\sim10\ {\rm g\ cm^{-2}}$ and zero at $r\sim110$ au and 265 au, respectively, and the profile linearly decreases as a function of radius, we can estimate the gas mass beyond $r\sim110$ au to be $\sim0.09\ {\rm M_{\odot}}$ and the total gas mass to be $\sim0.1-0.2\ {\rm M_{\odot}}$.}
We note that this value may be an upper limit as the gas surface density might decrease more steeply.
Therefore, we can roughly estimate the total gas mass to be $0.05-0.2\ {\rm M_{\odot}}$ with the lower and upper limits.

This overlaps the mass estimates of other samples from the kinematical analysis \citep{Longarini_exoALMA} and stands on the upper sides of the values from the molecular line analysis \citep{Trapman_exoALMA}.
If the actual mass is closer to the upper limit, the J1604 disk is one of the most massive disks in the sample.

In this study, the CO/${\rm H_2}$ ratio in the ring of the J1604 disk is found to be similar to or higher than the ISM value.
Among the samples, T-Tauri disks are found to have typically 3-30 $\times$ depleted in CO compared to the ISM \citep{Trapman_exoALMA, Rosotti_exoALMA}, which is different from the J1604 disk.
This might imply that there is radial variation in the CO/${\rm H_2}$ ratio; in the inner region, CO/${\rm H_2}$ is high due to thermal desorption and additional supply from radially drifted ice, while it may be depleted in the outer region \citep[e.g.,][]{zhan19, zhan21}.

\subsection{Caveats}
{ In our analysis, we assumed that { CO molecules are present} in the gas phase even in the disk midplane and that the CO/H$_2$ ratio is vertically constant.
If the CO molecules were frozen out near the midplane, the derived gas surface density would be underestimated.
}

Regarding the gas-to-dust ratio, it should be noted that there might be a large uncertainty in the dust opacity model.
For instance, the {\tt porous} dust model with the maximum dust size of 1 mm in \citet{birn18} has a $\sim 6 \times$ smaller absorption opacity at 3 mm wavelength.
Therefore, if we adopted the {\tt porous} dust model, the gas-to-dust surface density ratio at the ring peak would become { $10-70$}.
This value satisfies the favorable condition for streaming instability \citep{youd05}.
On the other hand, it has been recently proposed that higher absorption opacity models such as \citet{ricc10} better match observations \citep{delu24}.
In this case, the gas-to-dust ratio would become even larger than the results based on the DSHARP model.

Furthermore, we note that the spatial resolution of our observations is limited.
Therefore, the width of the gas surface density bump could be narrower in reality and caution needs to be exercised when comparing with the dust ring.
We note, however, that the gas ring is spatially resolved { at our resolution}, which means that this effect is not very significant.

In the CO spectra fitting, we assumed that the stacked spectrum is identical to the intrinsic one without stacking.
{
However, these two could differ in reality due to the inclination effect, although we approximated this effect by adopting the effective velocity resolution.
}
This might result in a potential error in the atmospheric temperature (Appendix \ref{app:temp}) which is effectively constrained from the narrow line core.
However, we emphasize that the atmospheric temperature is almost independent of the gas surface density (Appendix \ref{app:corner}).

\section{Summary} \label{sec:sum}
{ Thanks to the high-sensitivity observations of the exoALMA Large program, we detected pressure-broadened $\rm ^{12}CO$ line wings in the J1604 disk.
This is the second detection of pressure-broadening in protoplanetary disks, opening up further application of the method demonstrated by \citet{yosh22} and enabling us to robustly derive the gas surface density profile.
Notably, this method can achieve much higher accuracy by at least a factor of $\sim30$ compared to a conventional method that utilizes a CO isotopologue line intensity.
We modeled the $\rm ^{12}CO$ emission line by solving the radiative transfer equation and fitted it to the azimuthally averaged spectra.}
{ Based on the analysis, we conclude the following:

\begin{itemize}
\item We derived the gas surface density and midplane pressure profiles by using the pressure broadening effect. The pressure profile has a clear peak at the dust ring radius ($\sim 82\ {\rm au}$). This is direct evidence of radial dust trapping at the gas pressure maximum.

\item Assuming that the disk is gravitationally stable and the CO/H$_2$ ratio is within $10^{-3}-10^{-6}$, the peak gas surface density is constrained to be $18-44\ {\rm g\ cm^{-2}}$. The disk gas mass within $50<r<110$ au and within the ring are estimated to be $0.02-0.06\ {M_\odot}$ and $\sim 0.05-0.1\ {M_\odot}$, respectively, suggesting that there is a sufficient mass for planet formation to occur. The estimated mass range is comparable with the other disks in the exoALMA sample.

\item The gas surface density profile exhibits a drop at $r\sim60$ au.
If we attribute this to a planetary-induced gap, we can estimate the planetary mass to be $0.1-2\ M_{\rm jup}$.

\item We analyzed the multi-band continuum observations with a simple parametric model and derived the dust surface density profile.
Comparing the gas and dust surface densities, we find that the ring is not very enriched in dust.
This suggests that there is significant hidden dust mass and/or less efficient dust trapping.

\item From the dust ring width compared to the gas ring width, we constrained the turbulence strength in the ring to be $\alpha_{\rm turb} \sim 2\times10^{-4}$, which is consistent with previous studies in other disks.

\item Assuming that the maximum dust size is regulated by the turbulence-induced collisional fragmentation, the fragmentation threshold velocity $v_{\rm frag}$ is estimated to be $\sim1\ {\rm m\ s^{-1}}$, which suggests that the dust grains are fragile.

\end{itemize}

We note that the CO/H$_2$ ratio remains uncertain in this study. However, we can eliminate this uncertainty and even determine the CO/H$_2$ ratio if the CO optical depth is known.
Observations of optically thin CO isotopologue lines are required.
The methodology used in this Letter should be applicable to other disks with similar characteristics (such as inclination, disk mass, and temperature), providing a new window for high-precision studies of the physical properties of planet-forming environments.
}


\section*{Acknowledgments}
We would like to thank the anonymous referee for helpful comments.
This paper makes use of the following ALMA data: ADS/JAO.ALMA \# 2013.1.01020.S, 2015.1.00819.S, 2015.1.00964.S, 2017.A.01255.S, 2021.1.01123.L. ALMA is a partnership of ESO (representing its member states), NSF (USA) and NINS (Japan), together with NRC (Canada), MOST and ASIAA (Taiwan), and KASI (Republic of Korea), in cooperation with the Republic of Chile. The Joint ALMA Observatory is operated by ESO, AUI/NRAO and NAOJ.
The National Radio Astronomy Observatory is a facility of the National Science Foundation operated under cooperative agreement by Associated Universities, Inc.
TCY is supported by Grant-in-Aid for JSPS Fellows, JP23KJ1008.
S.F. is funded by the European Union (ERC, UNVEIL, 101076613).
Views and opinions expressed are however those of the author(s) only and do not necessarily reflect those of the European Union or the European Research Council. 
Neither the European Union nor the granting authority can be held responsible for them.
S.F. acknowledges financial contribution from PRIN-MUR 2022YP5ACE. JS has received funding from the European Research Council (ERC) under the European Union’s Horizon 2020 research and innovation programme (PROTOPLANETS, grant agreement No. 101002188). Computations have been done on the 'Mesocentre SIGAMM' machine hosted by Observatoire de la Côte d’Azur.
MM is supported by a Grant-in-Aid from the Japan Society for the Promotion of Science (KAKENHI: No. JP18H05441). JB acknowledges support from NASA XRP grant No. 80NSSC23K1312.
MB has received funding from the European Research Council (ERC) under the European Union’s Horizon 2020 research and innovation program (PROTOPLANETS, grant agreement No. 101002188).
PC acknowledges support by the Italian Ministero dell'Istruzione, Universit\`a e Ricerca through the grant Progetti Premiali 2012 – iALMA (CUP C52I13000140001) and by the ANID BASAL project FB210003.
DF has received funding from the European Research Council (ERC) under the European Union’s Horizon 2020 research and innovation program (PROTOPLANETS, grant agreement No. 101002188).
MF is supported by a Grant-in-Aid from the Japan Society for the Promotion of Science (KAKENHI: No. JP22H01274).
CH acknowledges support from NSF AAG grant No. 2407679. JDI acknowledges support from an STFC Ernest Rutherford Fellowship (ST/W004119/1) and a University Academic Fellowship from the University of Leeds.
Support for AFI was provided by NASA through the NASA Hubble Fellowship grant No. HST-HF2-51532.001-A awarded by the Space Telescope Science Institute, which is operated by the Association of Universities for Research in Astronomy, Inc., for NASA, under contract NAS5-26555.
CL has received funding from the European Union's Horizon 2020 research and innovation program under the Marie Sklodowska-Curie grant agreement No.823823 (DUSTBUSTERS) and by the UK Science and Technology research Council (STFC) via the consolidated grant ${\rm ST/W000997/1}$.
C.P. acknowledges Australian Research Council funding via FT170100040, DP18010423, DP220103767, and DP240103290.
D.P. acknowledges Australian Research Council funding via DP18010423, DP220103767, and DP240103290.
GR acknowledges funding from the Fondazione Cariplo, grant no. 2022-1217, and the European Research Council (ERC) under the European Union’s Horizon Europe Research \& Innovation Programme under grant agreement no. 101039651 (DiscEvol).
Views and opinions expressed are however those of the author(s) only, and do not necessarily reflect those of the European Union or the European Research Council Executive Agency.
Neither the European Union nor the granting authority can be held responsible for them.
H.-W.Y.\ acknowledges support from National Science and Technology Council (NSTC) in Taiwan through grant NSTC 113-2112-M-001-035- and from the Academia Sinica Career Development Award (AS-CDA-111-M03).
GWF acknowledges support from the European Research Council (ERC) under the European Union Horizon 2020 research and innovation program (Grant agreement no. 815559 (MHDiscs)). GWF was granted access to the HPC resources of IDRIS under the allocation A0120402231 made by GENCI.
AJW has received funding from the European Union’s Horizon 2020 research and innovation programme under the Marie Skłodowska-Curie grant agreement No 101104656. Support for BZ was provided by The Brinson Foundation.


%

\vspace{5mm}
\facilities{ALMA}


\software{astropy \citep{astropy:2013, astropy:2018}, emcee \citep{fore13}, eddy \citep{eddy}}



\bibliography{sample631}{}
\bibliographystyle{aasjournal}

\appendix

\section{Data reduction of the Band 3 image}
\label{sec:b3im}
We obtained the visibility data from the ALMA science archive (\# 2015.1.00819.S; PI. L. Ricci).
The J1604 disk was observed on Jan. 20, 2016 in Band 3 with 41 antennas.
The baseline lengths range from 15 m to 331 m, and the on-source integration time was 2.8 min.
All the spectral windows are used in the Time Division Mode.

We first ran the calibration script provided by the observatory using the Common Astronomical Software Application \citep[CASA; ][]{CASA} pipeline version 4.5.2.
CASA modular version 6.5.5 was used throughout the remaining analysis.
We split out only the data observing the J1604 disk from the calibrated dataset.
We then CLEANed the data and performed iterative self-calibration.
Since the relative bandwidth with respect to the observing frequency is large, we adopted the MTMFS mode for CLEAN. 
The Briggs weighting with robust $=$0.5 was applied.
Three rounds of phase self-calibration with solution intervals of the durations of the observation, 90s, and 30s were applied.
The final RMS noise level is measured to be 0.05 mJy, and the resultant beam size is $2\farcs7 \times 1\farcs7$.

\section{Multi-band continuum modeling results} \label{sec:cont_comparison}
\begin{figure}[hbtp]
    \epsscale{1.2}
    \plotone{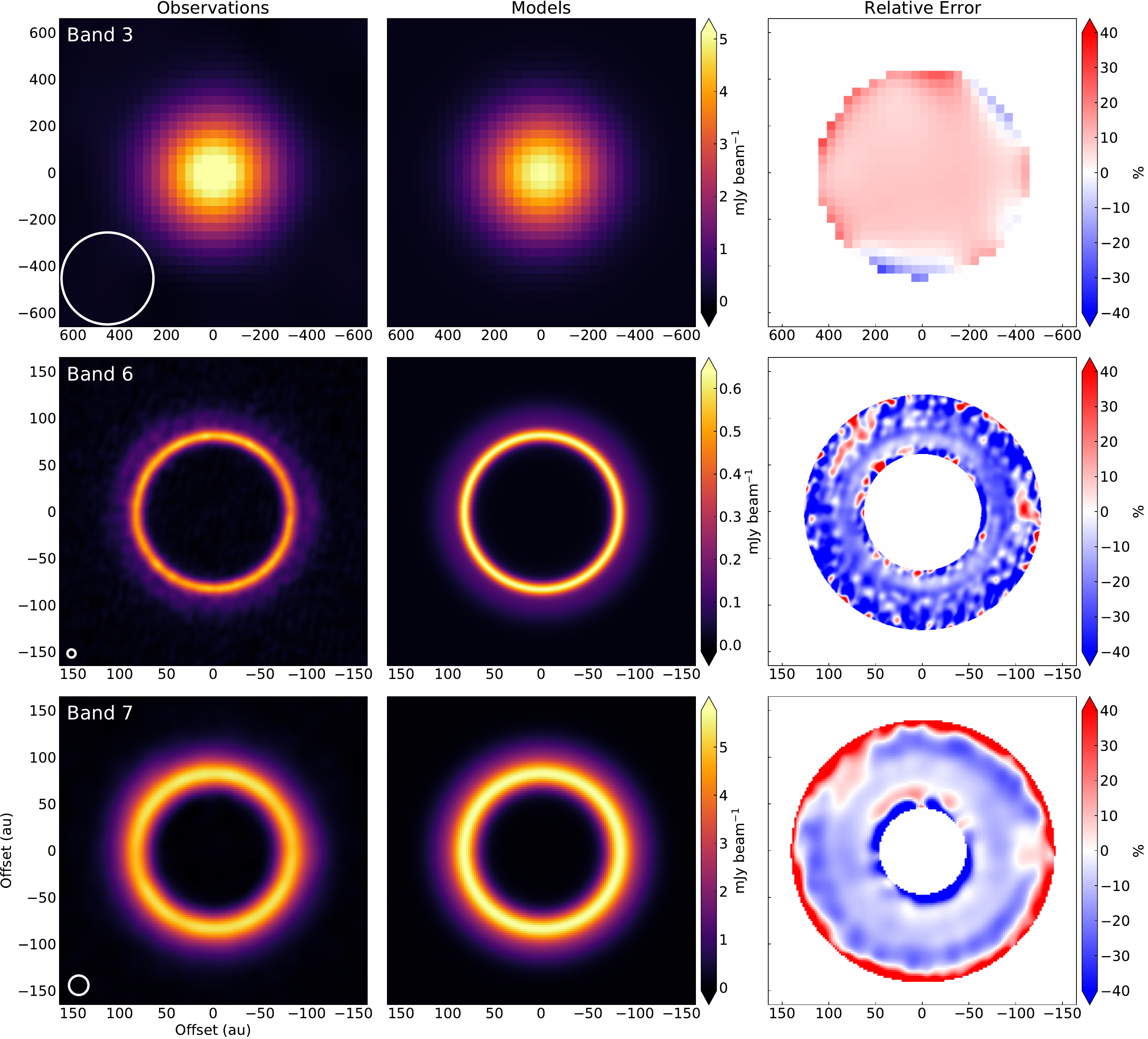}
    \caption{ Observed continuum images, best-fit model images, and relative error between the observations and model images. The white ellipses on the left column indicate the observational beam size. The error maps are masked by the $3\sigma$ noise level measured in the observation images. }
    \label{fig:cont}
\end{figure}

We compare the observed and modeled continuum images in Figure \ref{fig:cont}.
Since the absolute flux uncertainty dominates the spatial noise, we also plot the relative error between the observed and modeled images.

\section{CO line stacking} \label{sec:azimuthal}

\begin{figure}[hbtp]
    \epsscale{1.2}
    \plotone{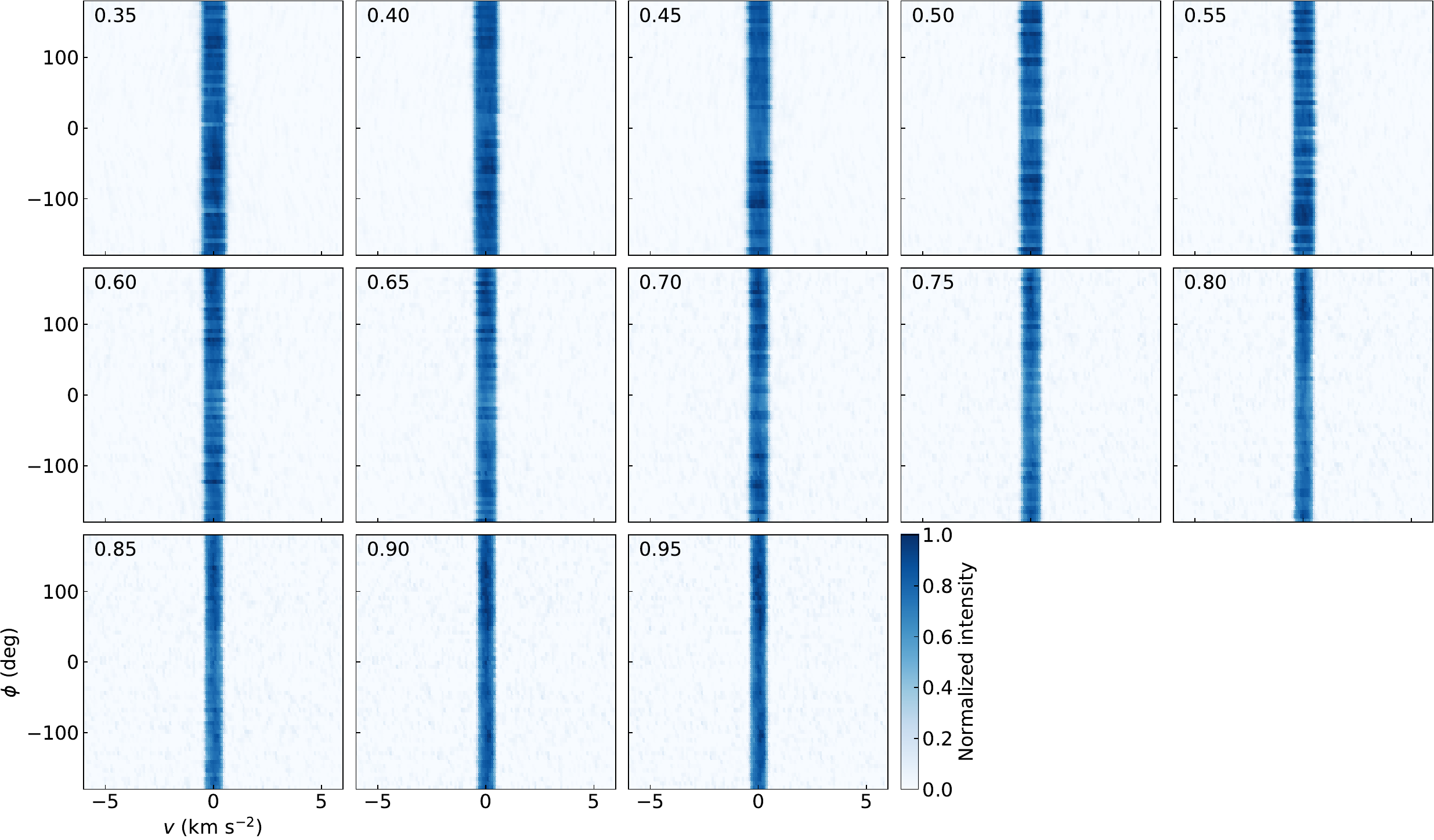}
    \caption{ Shifted spectra for each annulus. In each panel, the spectra are plotted as a function of velocity from the line center and azimuthal angle in the disk coordinate. The numbers in the upper left corner indicate the radius of the annulus in arcseconds.  }
    \label{fig:stacking}
\end{figure}
In Figure \ref{fig:stacking}, we present shifted spectra for each annulus as a function of velocity from the line center and azimuthal angle.

\begin{figure}[hbtp]
    \epsscale{1.2}
    \plotone{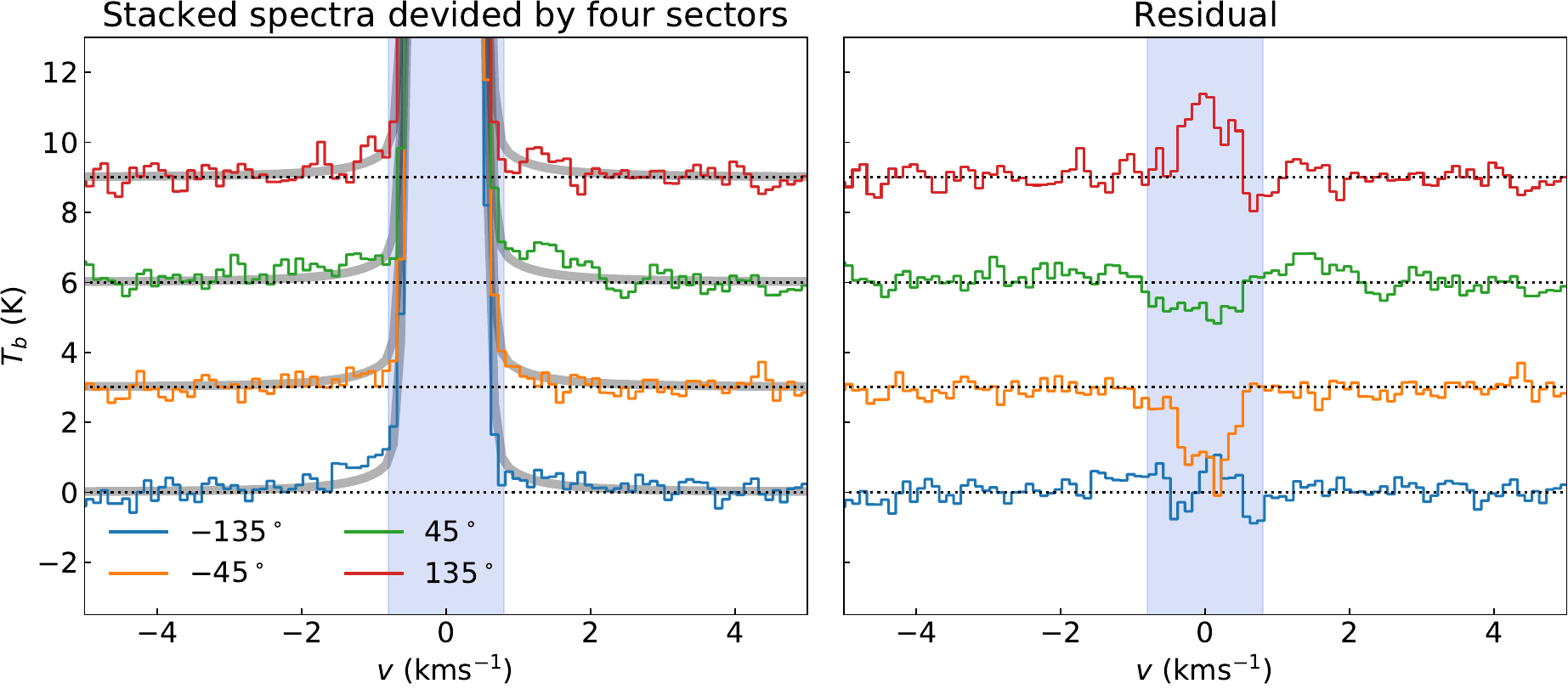}
    \caption{ (left) Radially-averaged azimuthally-stacked spectra at $r=0\farcs4-0\farcs8$. The stacked azimuthal range is limited to 1/4 of an annulus and the center azimuthal angle is shown in the legend. The grey solid lines indicate the fiducial model spectrum. (right) Residual of observed spectra after subtracting the fiducial model. Note that the spectra are vertically shifted for visibility purposes. The blue shaded area indicates the optically thick line cores.  }
    \label{fig:sectors}
\end{figure}
Figure \ref{fig:sectors} shows radially-averaged azimuthally-stacked spectra at $r=0\farcs4-0\farcs8$. Here, the azimuthal stacking is performed only over one-fourth sector of an annulus.
In the right panel, we show the residual of observed spectra after subtracting the fiducial model. There is no significant signal in the wing region (outside of the blue-shaded range) of the residual.
In the line core, some residuals remain. This is due to the azimuthal asymmetry of the peak intensity, which might be caused by the shadow cast by the center region \citep{stad23}.

\begin{figure}[hbtp]
    \epsscale{1.2}
    \plotone{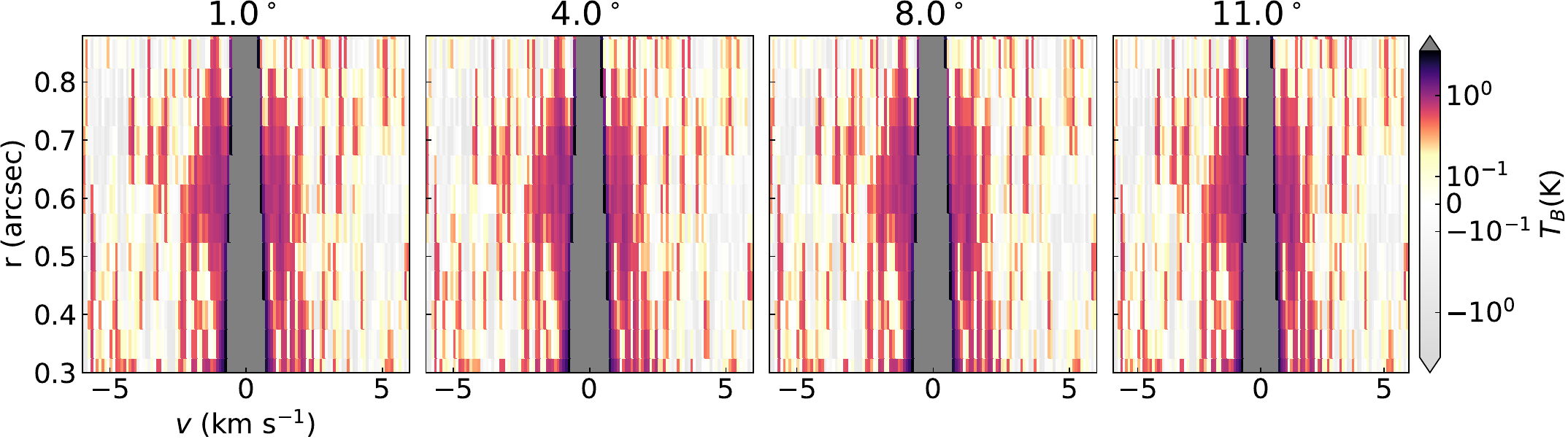}
    \caption{ Teardrop plots for different inclinations. The wing feature persists despite $5\arcdeg$ differences in the input inclination. { Unlike Figure \ref{fig:teardrop}, the velocity convolution is not applied for completeness. }}
    \label{fig:diff_incl}
\end{figure}
When we stacked the spectra, we fixed the disk inclination angle to $6\fdg0$ according to \citet{stad23} and fitted other geometrical parameters (Section \ref{sec:stacked}).
To test if the wing feature persists for different possible inclination angles, we changed the inclination to $1\fdg0, 4\fdg0, 8\fdg0$ and $11\fdg0$ and generated the teardrop plots again (Figure \ref{fig:diff_incl}).
Here, we fixed other parameters such as the position angle to the fiducial values used in the main text.
As a result, no significant difference is observed, which means that our analysis is robust for the uncertainty in the geometrical parameters.

\section{Corner plots}
\label{app:corner}
\begin{figure}[hbtp]
    \epsscale{1.0}
    \plotone{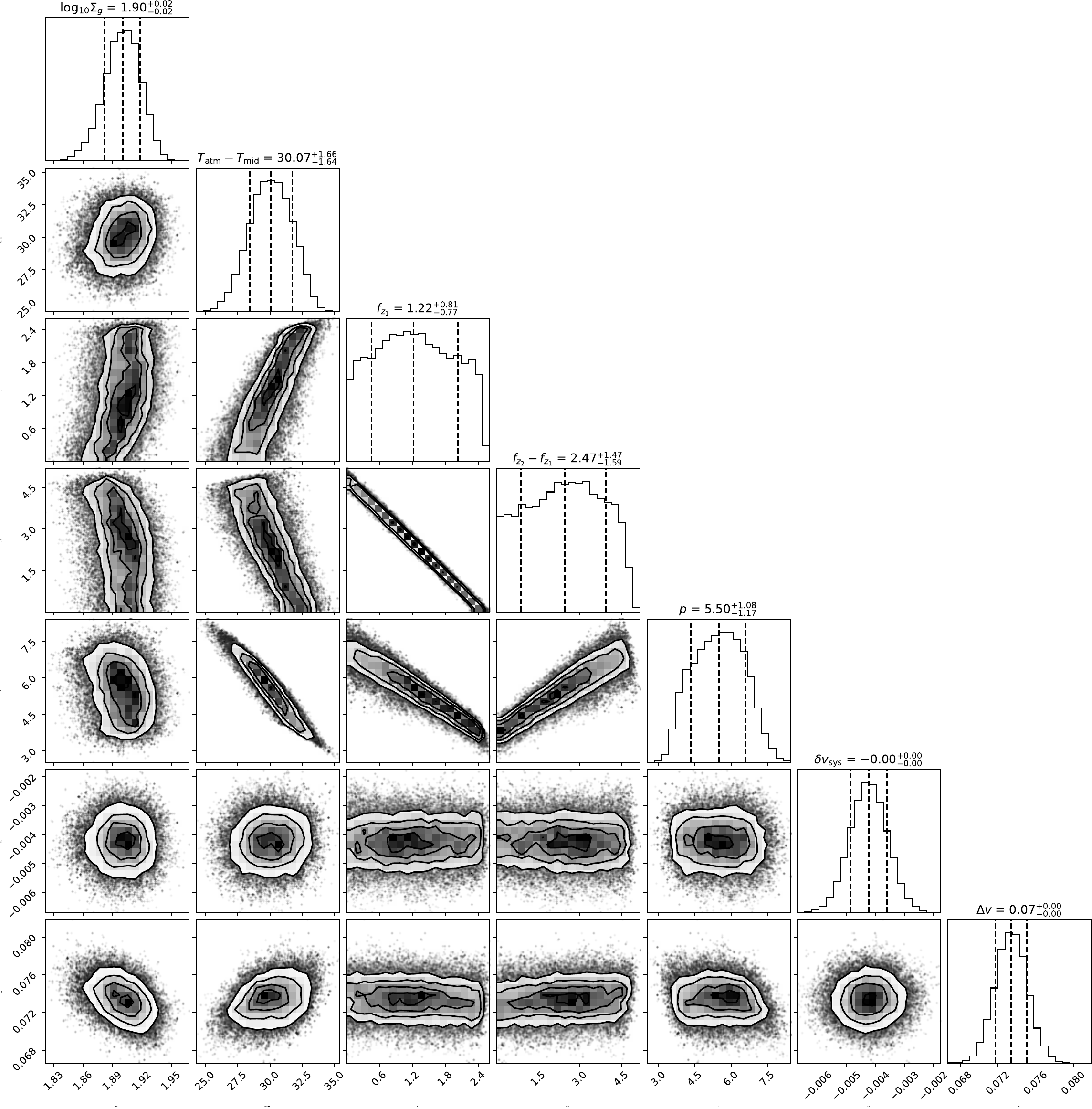}
    \caption{ Corner plot of the CO line fitting at $r = 0\farcs6$. }
    \label{fig:corner}
\end{figure}

A representative marginal posterior distribution of the CO line fitting at $r = 0\farcs6$ is presented in Figure \ref{fig:corner}.

\section{The case of the Gaussian profile} \label{sec:gaussian}

\begin{figure}[hbtp]
    \epsscale{1.2}
    \plotone{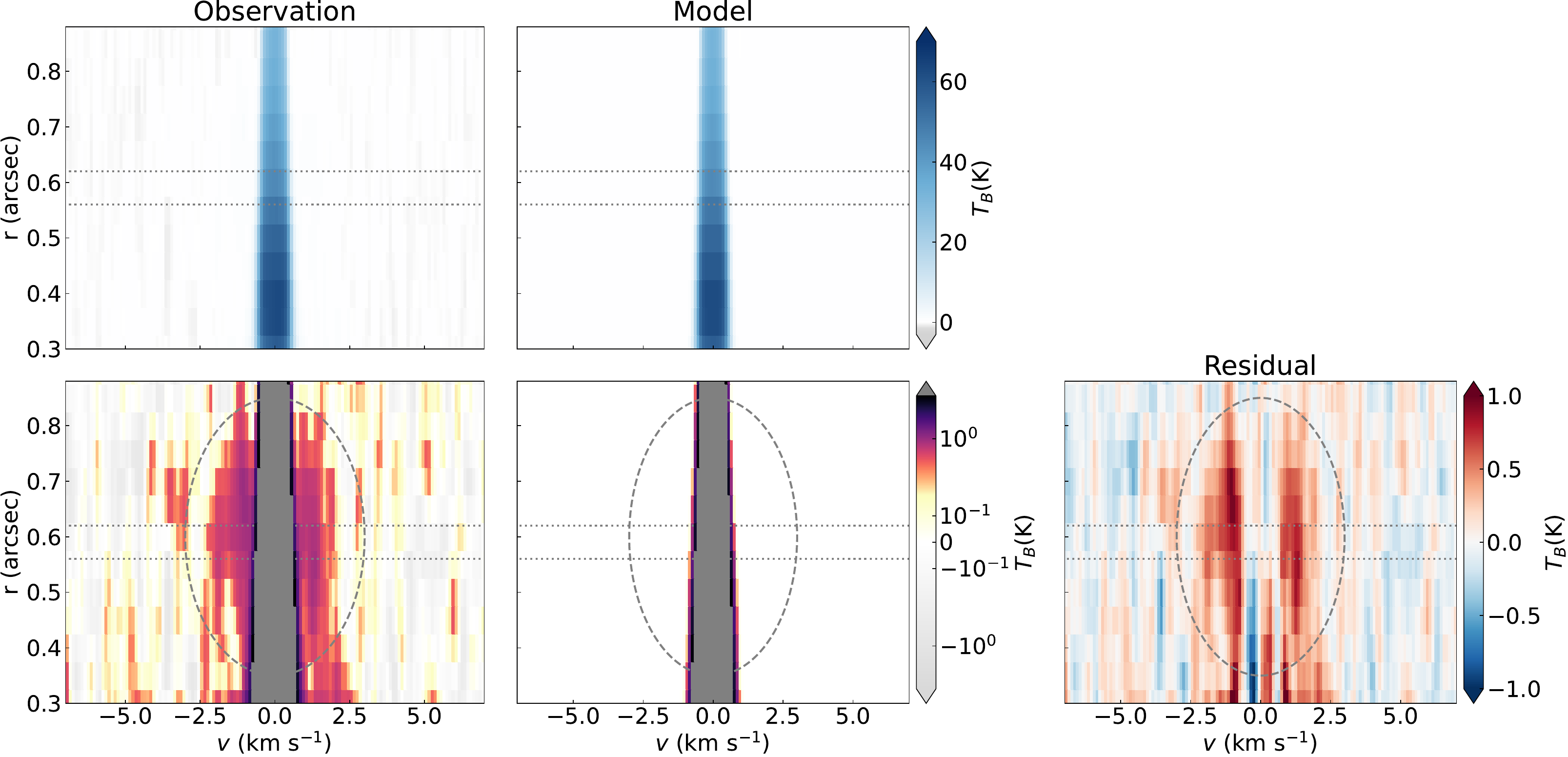}
    \caption{ The same as Figure \ref{fig:teardrop} but the Gaussian function is adopted for the opacity profile.}
    \label{fig:gaussian}
\end{figure}
Figure \ref{fig:gaussian} indicates the results when we adopt the Gaussian function as an opacity profile instead of the Voigt function.
There is a significant residual in the line wing region in contrast with the case of the Voigt function.

\section{Emitting layer} \label{app:layer}
\begin{figure}[hbtp]
    \epsscale{1.1}
    \plotone{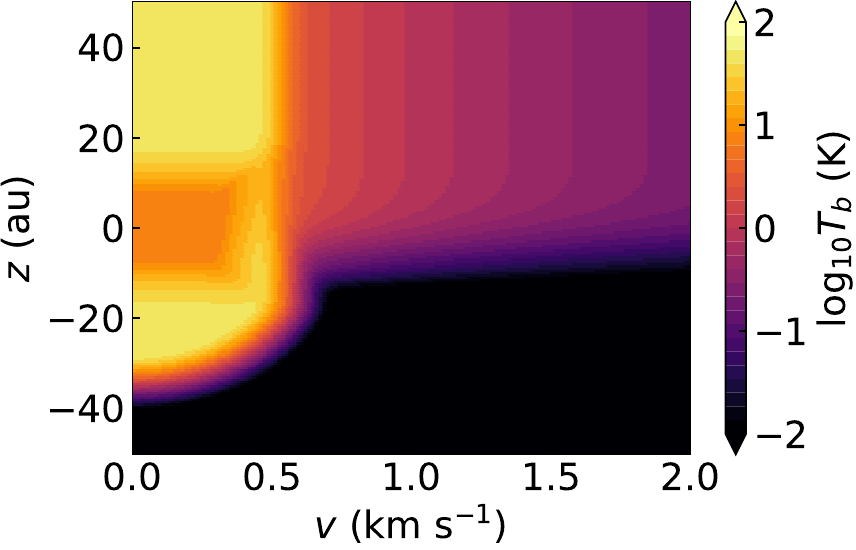}
    \caption{ Local specific intensity field along the disk vertical axis in a unit of brightness temperature. }
    \label{fig:vzTb}
\end{figure}
Figure \ref{fig:vzTb} shows the local specific intensity field as a function of $(z, v)$ at $r=80$ au in the best-fit model.
While the emission arises from the atmosphere at the line core ($v < 0.5\ \kms$), the emission in the wing ($v >0.5\ \kms$) originates from the region near the midplane where most of the gas inhabits. This is because the wings are optically thin.

\section{Temperature structure}\label{app:temp}

\begin{figure}[hbtp]
    \epsscale{0.9}
    \plotone{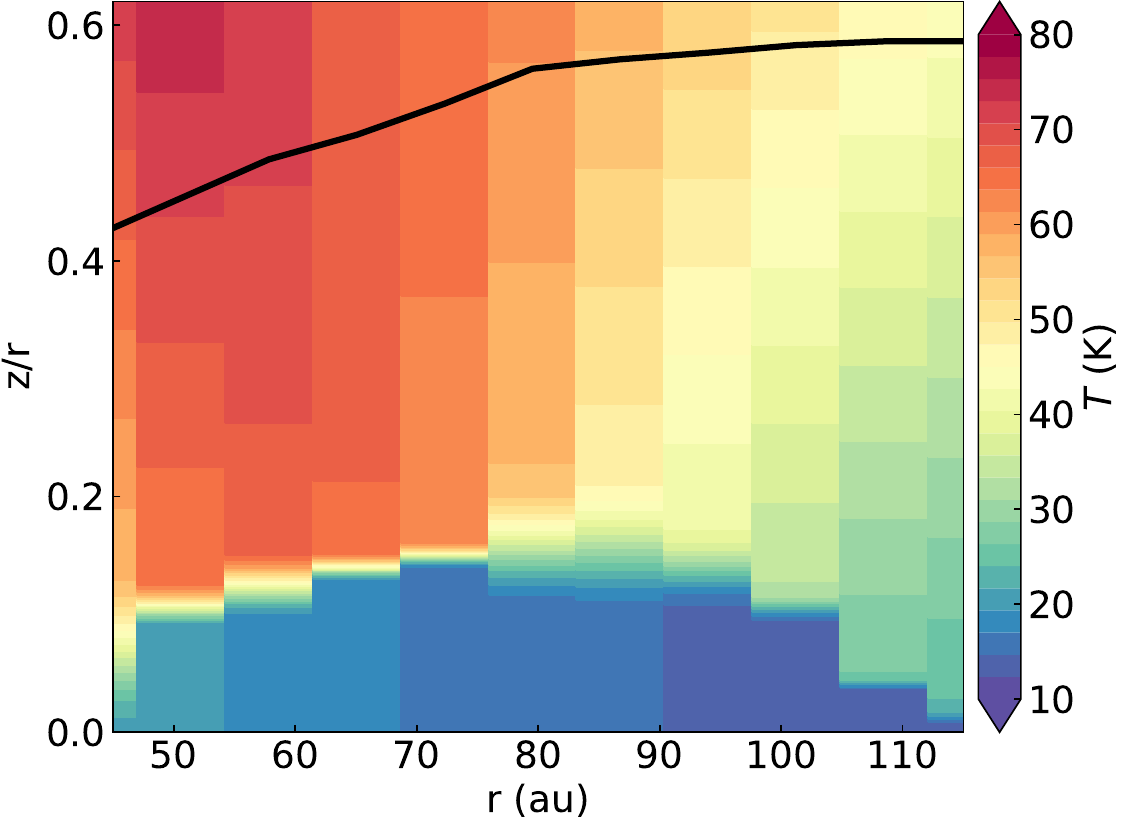}
    \caption{ Two-dimensional temperature structure constructed by using the best-fit parameters of the spectral model fitting.  }
    \label{fig:T}
\end{figure}
By fitting the parametric temperature model, we reconstructed the 2D temperature structure as shown in Figure \ref{fig:T}.
We plot the kinetic temperature traced by the optically thick CO line center in the black solid line in Figure \ref{fig:T}.
It is shown that the emitting layer of the CO line center is $z/r\sim0.4-0.6$, which is on higher sides among measured values in other disks by channel map analysis and roughly consistent \citep[e.g.,][]{pint18, law21, law22}.
However, since we rely on the parametric model, there may be some uncertainty when comparing with more model-independent measurements.
Even though, the gas surface density measurements do not depend on the vertical temperature structure as the line wing emission is essentially optically thin, which is clear in the corner plot (the upper left panel in Figure \ref{fig:corner}).

\section{The ${\rm C^{18}O}\ {J=2-1}$ line} \label{app:c18o}
{
The detection of pressure-broadened line wings in the $\rm ^{12}CO$ line motivates us to check less optically thick isotopologue lines.
As discussed in Section \ref{sec:gas_res}, an optically thin isotopologue line such as $\rm ^{13}C^{17}O$ and $\rm ^{13}C^{18}O$ can solve the degeneracy between the CO column density and the H$_2$ density.
However, there is no available observation of them.
Nonetheless, it would be worth confirming that more abundant isotopologues than $\rm ^{13}C^{17}O$ and $\rm ^{13}C^{18}O$ are detected and likely optically thick.
Indeed, \citet{Galloway_exoALMA} confirms that the ${\rm ^{13}CO}\ J=3-2$ line is highly optically thick.
The next less-abundant isotopologue is ${\rm C^{18}O}$.
\citet{dong17} and \citet{berg19} observed the ${\rm C^{18}O}\ {J=2-1}$ line.

We downloaded the datasets published by \citet{dong17} and \citet{berg19} from the ALMA Science Archive and ran pipeline calibration (IDs: \# 2013.1.01020.S, 2015.1.00964.S).
We refer the readers to the original literature for details of each observation.
We combined the measurement sets to achieve higher sensitivity.
After subtracting the continuum emission, we created an image cube of the ${\rm C^{18}O}\ {J=2-1}$ line using the CASA command {\tt tclean} with the robust parameter of 0.5 and channel width of $0.2\ {\rm km\ s^{-1}}$.
We then smoothed the image cube with {\tt imsmooth} and achieved a beam size of $0\farcs26 \times 0\farcs26$. The resultant rms noise level is $8.8\ {\rm mJy\ beam^{-1}}$ in one channel.
The integrated intensity and peak intensity maps are generated using the Python script {\tt bettermoment} \citep{teag18} by specifying the velocity channel of $4.0 < v < 5.2\ {\rm km\ s^{-1}}$.

\begin{figure}[hbtp]
    \epsscale{0.8}
    \plotone{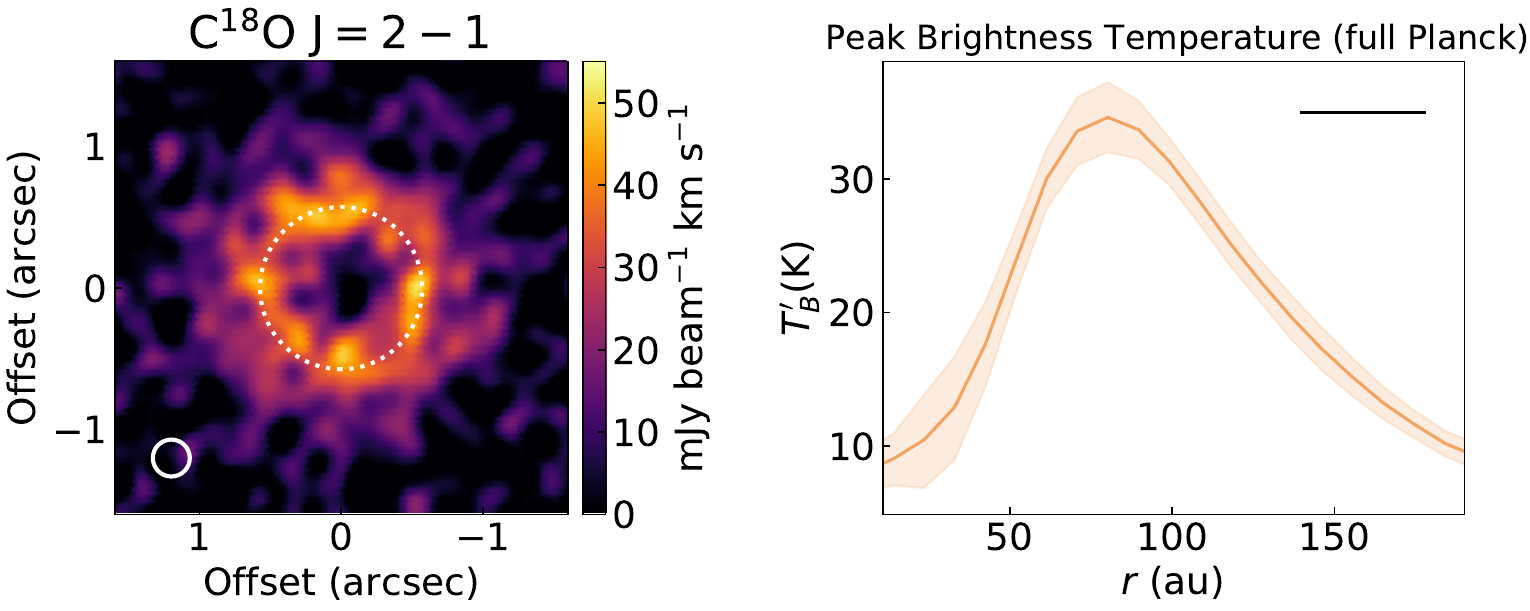}
    \caption{ { (left) Integrated intensity map of the ${\rm C^{18}O}\ {J=2-1}$ line. The white dotted circle indicates the dust ring location. The while solid circle shows the beam size. (right) Radial profile of the peak brightness temperature. The black bar indicates the beam size.} }
    \label{fig:c18o}
\end{figure}
The integrated intensity map in Figure \ref{fig:c18o} shows ring-like emission, which is consistent with \citet{dong17} and \citet{berg19}.
The integrated intensity near the ring is $\sim 40\ {\rm mJy\ beam^{-1}\ km\ s^{-1} }$.
In the right panel of Figure \ref{fig:c18o}, we present the radial profile of the peak brightness temperature using the full Planck function,
\begin{equation}
    T_{B}^{\prime} = \frac{h\nu}{k_B}\left [ \ln\left( \frac{2 h \nu^3}{ c^2 I_{\rm C^{18}O} } +1 \right) \right ]^{-1},
\end{equation}
where $\nu$ is the frequency of the transition and $I_{\rm C^{18}O}$ is the peak intensity.
The peak brightness temperature exceeds $30$ K around the ring at $r \sim 80$ au and reaches $\sim35$ K at the peak.

We found that the integrated intensity near the ring is exceptionally high as a ${\rm C^{18}O}\ J=2-1$ line among T Tauri disks when comparing similar radial locations ($r\sim80$ au). 
Indeed, this has been be already seen in \citet{berg19}.
With their spatial resolution ($\sim 0\farcs8$), the peak integrated intensity around $r\sim80$ au is $\sim 180\ {\rm mJy\ beam^{-1}\ km\ s^{-1}}$.
This is much higher than that in the other eight T Tauri disks in their sample but comparable with that in the Herbig disks around HD 163296 and MWC 480.
In addition, the MAPS project observed the same line in five disks with a better spatial resolution of $0\farcs15$ \citep{ober21, zhan21}.
If we scale the integrated intensity in our map to the $0\farcs15$ resolution, we obtain $\sim 13\ {\rm mJy\ beam^{-1}\ km\ s^{-1} }$.
This value is again much higher than similar regions in T Tauri disks, IM Lup, GM Aur, and AS 209 by a factor of $2-6$, but comparable with the Herbig disks (HD 163296 and MWC 480).

The peak brightness temperature of $\sim35$ K suggests that the ${\rm C^{18}O}\ J=2-1$ line is highly optically thick assuming that the midplane temperature is $\leq 35$ K.
We note that the limited spatial resolution could underestimate the brightness temperature, implying that the actual brightness temperature is even higher.
We can also compare it with the brightness temperature profiles of the same line in the MAPS sample \citep{law21}.
Among these sources, only the very inner region ($r<50$ au) of the Herbig disks (HD 163296 and MWC 480) show brightness temperatures higher than $\sim$30 K.


%
}




\end{document}